%% file: main.tex
\documentclass[final, shortpaper]{clv2025}

\jvol{vv}
\jnum{nn}
\jyear{2025}

\dochead{Short Paper}

\pageonefooter{Action editor: Wei Gao. Submission received: 26 May 2025; revised version received: 14 November 2025; accepted for publication: 23 November 2025.}

\usepackage{amsmath}
\usepackage{booktabs}

\usepackage{appendix}
\usepackage{chngcntr}

\usepackage{booktabs}

\usepackage{tabularray}
\UseTblrLibrary{booktabs}

\usepackage{tabularx}

\newcolumntype{Y}[1]{>{\centering\arraybackslash}X[#1]}

\usepackage{makecell}

\usepackage{multirow}

\usepackage{wrapfig}

\usepackage{subfig}

\newcommand{\misinfotblDataset}[2]{%
    \makecell[tc]{%
        {\footnotesize #1}\\
        {\tiny \cite{#2}}%
    }%
}

\newcommand{\misinfotblLabel}[1]{%
    {\footnotesize #1}
}

\newcommand{\misinfotblSize}[1]{%
    {\footnotesize #1}
}

\newcommand{\misinfotblDescription}[1]{%
    {\footnotesize #1}
}

\NewTblrTheme{misinfodatasets}{
  \DefTblrTemplate{firsthead}{default}{\UseTblrTemplate{caption}{default}}

  \DefTblrTemplate{firstfoot, middlefoot}{default}{
    \UseTblrTemplate{contfoot}{default}
  }

  \DefTblrTemplate{middlehead, lasthead}{default}{}

  \SetTblrStyle{caption-tag}{font=\small\bfseries}
  \DefTblrTemplate{caption-tag}{default}{
    \tablename\nobreakspace\thetable\par
  }

  \DefTblrTemplate{caption-sep}{default}{}

  \SetTblrStyle{caption-text}{font=\small\raggedright}
  \DefTblrTemplate{caption-text}{default}{
    \InsertTblrText{caption}\par\vskip-0.5\belowcaptionskip
  }

  \DefTblrTemplate{caption}{default}{
    \UseTblrTemplate{caption-tag}{default}
    \UseTblrTemplate{caption-sep}{default}
    \UseTblrTemplate{caption-text}{default}
  }

  \DefTblrTemplate{note-tag}{default}{\textsuperscript{\InsertTblrNoteTag}}
  \DefTblrTemplate{note-sep}{default}{\space}
  \DefTblrTemplate{note-text}{default}{{\small \InsertTblrNoteText}}

  \DefTblrTemplate{note}{default}{
    \MapTblrNotes{
    \noindent
    \UseTblrTemplate{note-tag}{default}
    \UseTblrTemplate{note-sep}{default}
    \UseTblrTemplate{note-text}{default}
    \par
    }
  }

}

\usepackage{microtype}
\usepackage{amssymb}
\usepackage{amsmath}

\usepackage{dsfont}

\usepackage{enumitem}

\usepackage[vskip=0.1em, leftmargin=10pt, rightmargin=10pt]{quoting}

\usepackage{siunitx}

\usepackage[dvipsnames]{xcolor}

\usepackage{layouts}

\hypersetup{
    pdfauthor={Ivo Verhoeven, Pushkar Mishra, Ekaterina Shutova},
    pdftitle={Yesterday's News: Multi-Dimensional Out-of-Distribution
Generalization of Misinformation Detection Models},
}

\usepackage{hyperref}

\newcommand{\reliable}{\textcolor{ForestGreen}{reliable$^{\boldsymbol{+}}$ }}
\newcommand{\unreliable}{\textcolor{BrickRed}{unreliable$^{\boldsymbol{-}}$ }}

\jinfo{Accepted for Publication}

\runningtitle{Yesterday's News}
\runningauthor{Verhoeven, Mishra, Shutova}

\begin{document}

\title{Yesterday's News: Benchmarking Multi-Dimensional Out-of-Distribution Generalization of Misinformation Detection Models}

\author{Ivo Verhoeven\thanks{Equal contribution}\thanks{Corresponding authors}$^{1}$, Pushkar Mishra$^{2}$, Ekaterina Shutova$^{1}$}

\affilblock{
  \affil{ILLC, University of Amsterdam\\\quad \email{i.o.verhoeven@uva.nl}}
  \affil{Meta AI, London}
}



\maketitle

\begin{abstract}
  This article introduces \texttt{misinfo-general}, a benchmark dataset for evaluating misinformation models' ability to perform out-of-distribution generalization. Misinformation changes rapidly, much more quickly than moderators can annotate at scale, resulting in a shift between the training and inference data distributions. As a result, misinformation detectors need to be able to perform out-of-distribution generalization, an attribute they currently lack. Our benchmark uses distant labelling to enable simulating covariate shifts in misinformation content. We identify time, event, topic, publisher, political bias, misinformation type as important axes for generalization, and we evaluate a common class of baseline models on each. Using article metadata, we show how this model fails desiderata, which is not necessarily obvious from classification metrics. Finally, we analyze properties of the data to ensure limited presence of modelling shortcuts. We make the dataset and accompanying code publicly available
  \footnote{\href{https://github.com/ioverho/misinfo-general}{\url{https://github.com/ioverho/misinfo-general}}}.
\end{abstract}


\section{Introduction}
\label{sec:introduction}

The field of misinformation detection aims to develop classification models that reliably moderate online content. Despite burgeoning academic interest \cite{wuMisinformationSocialMedia2019,zhouSurveyFakeNews2020} in a multitude of fields \cite{kruijverDisinformationLifecycleIntegrated2025}, and impressive classification results on existing datasets, mis- and disinformation content continues to propagate online and cause significant societal harm.

The rapid evolution of online content---which significantly outpaces model development cycles---partially explains this. News, and more generally all online content, is valued primarily for its novelty. It will often contain unseen entities, events and entity-event relationships. Further exacerbating issues is the fact that what does or does not constitute misinformation is constantly changing, and highly dependent on the perspective of the labeller \cite{yeeLimitsMachineLearning2025}. Verifying content manually using domain experts is prohibitively expensive, and typically requires context not available when news is emerging.

The misinformation datasets used for misinformation detector training, however, are collections of yesterday's news. These datasets typically have a narrow focus on particular events or misinformation forms, and are collected well after the fact. Consequently, state-of-the-art moderation systems lag behind the news landscape, and will encounter inference-time data distributions that have shifted away from the distribution of the training data. For misinformation detection to be successful at mitigating harms during deployment, and especially in situations with limited availability of social or historical context (i.e., when news is emerging), models will need to be robust to many forms of distribution shifts.

Currently, this property of Out-of-Distribution (OoD) generalization is lacking in many SoTA NLP models, and especially misinformation detection models. Performance significantly degrades when evaluated on unseen:
\begin{itemize}
  \item time periods \cite{bozarthBetterPerformanceEvaluation2020,horneRobustFakeNews2020, kochkinaEvaluatingGeneralisabilityNeural2023, stepanovaTemporalGeneralizabilityMultimodal2023},
  \item publishers \cite{rashkinTruthVaryingShades2017, zhouHiddenBiasesUnreliable2021a},
  \item events \cite{leeUnifyingMisinformationDetection2021, chengVRoCVariationalAutoencoderaided2020, dingMetaDetectorMetaEvent2022,wuProbingSpuriousCorrelations2022},
  \item topics \cite{przybylaCapturingStyleFake2020},
  \item domains \cite{hoyExploringGeneralisabilityFake2022, kochkinaEvaluatingGeneralisabilityNeural2023, verhoevenMoreRealisticEvaluation2024},
  \item cultures, or languages \cite{horneAllGoodActors2020, chuCrossLanguageFakeNews2021, ozcelikCrossLingualTransferLearning2023}.
\end{itemize}
We primarily attribute this to the present state of misinformation datasets. While plentiful, these are often small, collected over a short time span, centered around specific events, biased towards popular content, or contain a too homogenous set of publishers. These properties are generally believed to be detrimental to the generalization capabilities of modern NLP models, which require large, diverse pre-training datasets, especially when text or labels are noisy.

Creating a dataset that \textit{does} enforce OoD generalization, however, is not easy. Given the expense involved in collecting these datasets, prior attempts at doing so have invariably had to make trade-offs in size, diversity, or label fidelity (see Section \ref{sec:survey_misinformation_datasets}). As a result, these datasets are not representative of the misinformation landscape, and evaluation with such datasets will overestimate model performance during deployment \cite{aimeurFakeNewsDisinformation2023, xiaoChallengesMachineLearning2024, kunturFakeNewsDetection2024}. Generating high-quality misinformation labels for a realistically sized, naturalistic dataset remains intractable due to the cost of domain experts and the inherent subjectivity present in online content. This article will not solve this problem. Instead, we focus on more accurately estimating a model's robustness to expected distributional shifts.

Specifically, we present \texttt{misinfo-general}, a dataset meant for testing the generalization performance of automated misinformation detectors holistically. We do so by processing a distantly labelled series of corpora intended for publisher reliability labelling. While this introduces noise into the labels, we argue that the scale and diversity of the data make it useful for \textit{generalizability} evaluation. To mitigate said noise, we perform extensive pre-processing of the data (Section \ref{sec:misinfo-general}), and post-hoc testing of dataset properties (Section \ref{sec:dataset_analysis}). This ensures a balance of article quantity and label quality, providing one with rich metadata for a heterogeneous set of publishers, across a long time-span, covering a multitude of events and topics.

To showcase the utility of such a benchmark, we identify and operationalize six generalization axes---(1) time, (2) specific events, (3) topics, (4) publisher style, (5) political bias and (6) misinformation type. We then train a simple, yet representative baseline model. We find that generalization to different classes of publishers is particularly challenging, whereas within-publisher variation across years is smaller than expected. Using the metadata available to us, we provide additional analysis of publisher-level determinants of performance, and find some undesirable model behaviors not discussed in prior literature: less frequent publishers see degraded performance, and models treat different political biases differently. Juxtaposed to these results, we also find some initial evidence that the scale and diversity of this dataset can benefit model generalization ability when trained on.


\section{Related Work}
\label{sec:related_works}

\subsection{Generalizable Misinformation Detection}
\label{sec:related_works_generalizable_misinformation}

Generalization abilities of misinformation classifiers have been tested in many settings, at smaller scales. \citet{horneRobustFakeNews2020} found that performance degrades quickly when evaluating on future events, which \citet{bozarthBetterPerformanceEvaluation2020} corroborate and extend to changes in domain. The same issues have also been reported in misinformation detection in other modalities \cite{stepanovaTemporalGeneralizabilityMultimodal2023, verhoevenMoreRealisticEvaluation2024}. \citet{zhouHiddenBiasesUnreliable2021a} find that models tend to overfit to publisher idiosyncrasies more than article content, especially in publisher-level annotated datasets.

Results on existing benchmark datasets are generally not indicative of downstream performance. \citet{kochkinaEvaluatingGeneralisabilityNeural2023} found that performance within one dataset vastly overestimates performance on other datasets or time spans. Even when controlling for the time period or topic, \citet{hoyExploringGeneralisabilityFake2022} found that models overfit to the training dataset and perform worse on similar but unseen datasets. In a recent systematic review of the literature, \citet{xiaoChallengesMachineLearning2024} come to the conclusion that:
\begin{quote}
  \ldots detection tasks are often meaningfully distinct from the challenges that online services actually face. Datasets and model evaluation are often non-representative of real-world contexts, and evaluation frequently is not independent of model training (p. 1)
\end{quote}
This sentiment matches the earlier discussion in \citet{aimeurFakeNewsDisinformation2023}; current misinformation benchmarks and evaluation setups can yield deceptively high performance scores.

Despite this paucity in benchmarks and labels, there has been some interest in developing generalizable or adaptive misinformation detection techniques. This has been attempted through weak supervision \cite{shuLeveragingMultiSourceWeak2020}, multitask training \cite{leeUnifyingMisinformationDetection2021}, utilizing external agents \cite{dingMetaDetectorMetaEvent2022, mosallanezhadDomainAdaptiveFake2022}, data resampling or active learning \cite{huLearnEvolveFuture2023}, adversarial learning \cite{linDetectRumorsMicroblog2022a}, or gradient-based meta-learning \cite{zhangLearningDetectFewShotFewClue2021, yueContrastiveDomainAdaptation2022}. While these research directions are promising, their utility for out-of-distribution misinformation detection has not been sufficiently tested on large, diverse benchmark data.

\subsubsection{Synthetic Distribution Shifts}
\label{sec:related_works_synthetic_ditribution_shifts}

This article focuses on exploring the robustness of automated misinformation detectors to natural distribution shifts, i.e., those one might expect to occur during transfer from training to deployment-time inference.

A related strand of research is the analysis of model performance under \textit{synthetic} distribution shifts. These techniques can avoid the cost of collecting and extracting misinformation data, while elucidating model behavior under covariate shifts and adversarial attacks.

In general, misinformation classifiers have been found to be fragile against adversarial attacks \cite{zhouFakeNewsDetection2019, koendersHowVulnerableAre2021}. \citet{przybylaVerifyingRobustnessAutomatic2024} found larger LMs were more fragile to data augmentation techniques that minimize semantic distance, while maximizing performance degradation. Despite this, those same LMs were successful in generating adversarial examples \cite{przybylaAttackingMisinformationDetection2025}. On the other hand, several recent works find that incorporating adversarial data augmentation techniques during training \cite{smithMitigatingAttacksFake2021, ahmedEffectTextAugmentation2024} can boost robustness. In extreme cases, LLM-generated misinformation is used as a proxy for sampled misinformation data in misinformation detector evaluation \cite{lucasFightingFireFire2023}, which theoretically should allow for fine-grained control over distribution shifts being tested.

\subsection{Publisher Reliability Estimation}
\label{sec:related_works_publisher_reliability}

A related field to misinformation classification, especially when utilizing publisher-level labels, is publisher reliability estimation. Instead of yielding article level moderation decisions, a publisher reliability model uses the content of one or many articles from one publisher to yield a reliability estimate of the publisher as a whole. This is a relatively well-studied problem. At present, this is usually achieved through a mix of content-based \cite{rashkinTruthVaryingShades2017,bianchiEvaluatingTrustworthinessOnline2024} and metadata features \cite{balyPredictingFactualityReporting2018, balyIntegratingStanceDetection2018,balyMultiTaskOrdinalRegression2019,balyWeCanDetect2020,balyWhatWasWritten2020, nakovSurveyPredictingFactuality2024}.

Relative to article-level misinformation classifiers, publisher-level classification can greatly reduce the computational cost needed for classification \cite{burdissoReliabilityEstimationNews2024}. However, this typically involves incorporating additional historical context, world-knowledge \cite{yangAccuracyPoliticalBias2025} or social context \cite{pratelliUnveilingNewsPublishers2024}. This can make publisher reliability models \textit{transductive} instead of \textit{inductive} learners---moderation decisions come from specific prior experience rather than general rules.

This mimics how moderators or users might analyze the reliability of a publisher, potentially before ingesting the contents of a specific article. However, such approaches might fail in cases of where publishers are unknown, ambiguous or evolving. In those situations, moderation decisions at the article-level is necessary. While \texttt{misinfo-general} is suited for either approach, we focus on testing the generalization of inductive article-level classifiers. These models naturally provide classification in cases where limited context or prior experience is available, and are required to utilize (non-spurious) general rules.

\section{Biases in Misinformation Datasets}
\label{sec:survey_misinformation_datasets}

At risk of repetition: misinformation models' performance degrades quickly under covariate distribution shifts expected to occur during model deployment, an observation whose cause we attribute to the datasets they were trained on. Due to the exorbitant cost of acquiring high-fidelity misinformation labels, misinformation datasets tend not to reflect the true variance in online (misinformation) content.

To illustrate this, we analyze common properties of datasets specifically constructed for the development of misinformation detectors, by ways of an inexhaustive, yet representative survey of existing misinformation datasets. We provide an overview of these datasets in Appendix \ref{app:survey} Table \ref{tab:misinfo_datasets}. Furthermore, in this section, we (1) broadly categorize datasets into different labelling methods; (2) provide specific examples of how misinformation is collected and labelled; (3) discuss how these operationalizations can lead to biases in the datasets; (4) and finally, provide a discussion on the merits and demerits of publisher-level labelled datasets for the purposes of model generalization.

\subsection{Dataset Labelling Granularity}

Generally speaking, one can classify misinformation datasets into $3$ annotation schemes. Listed from most fine-grained to most coarse-grained:
{
\begin{enumerate}
  \item \textbf{Claim}: experts fact-check individual (but complete) statements in isolation. Claims are usually small spans sourced from larger documents or utterances
  \item \textbf{Article}: experts label the \textit{overall} veracity of entire documents. These can contain many claims, whose factuality need not be consistent with each other
  \item \textbf{Publisher}: experts label publishers for their propensity for factual reporting, based on historical records and prescribed authorial intent. These labels are often used as a proxy for finer-grained labels. The articles produced by publishers do not necessarily have the same label as the publisher
\end{enumerate}
}

The more fine-grained annotation methods yield high-quality labels, but can be prohibitively expensive to procure, or evaluate texts without the context those texts would naturally have. Furthermore, these labelling methods are typically forced to exclude unverifiable texts (e.g., highly subjective texts or opinions), despite these being prevalent in online discourse. On the other hand, the more coarse-grained annotation methods run the risk of introducing noise into the labels, by assuming consistency between finer-grained labels. For example, an article may contain many factual statements, but a single blatant lie. Since there are increasingly fewer units at each level, however, labels are far easier to procure.


\subsection{Survey of Misinformation Datasets}

In Appendix \ref{app:survey} Table \ref{tab:misinfo_datasets} we present various misinformation datasets, their labelling granularity, their size, and a description of how their data was sampled. In this subsection, we briefly expand on some common trends on how misinformation data and labels were sourced.

`Claim'-level annotations represent some of the oldest (\textsc{Lie Detector} \cite{mihalceaLieDetectorExplorations2009}) and largest (\textsc{CREDBANK} \cite{mitraCREDBANKLargeScaleSocial2015}) collections of misinformation text. The claims can be sourced from directly sampling social media (\textsc{CREDBANK} \cite{mitraCREDBANKLargeScaleSocial2015}) or sampling specific utterances flagged for review (\textsc{Liar} \cite{wangLiarLiarPants2017}, \textsc{PolitiFact-Oslo} \cite{poldverePolitiFactOsloCorpusNew2023}).

While labels sourced from domain experts are dominant, using lay people as a method of crowdsourcing for either data or label collection has also proven popular. For the former, as an example, articles are collected only if these were flagged by (trusted) users of social media sites (\textsc{Weibo15} \cite{maDetectingRumorsMicroblogs2016}, \textsc{Weibo17} \cite{jinMultimodalFusionRecurrent2017}, \textsc{WeChat} \cite{wangWeakSupervisionFake2020}). In some cases, lay volunteers were even used in the production of misinformation (\textsc{Lie Detector} \cite{mihalceaLieDetectorExplorations2009}, \textsc{FakeNewsAMT} \cite{perez-rosasAutomaticDetectionFake2018}).

The benefit of crowdsourcing is clear; especially at the article level, datasets that use expert annotations (\textsc{BuzzFeed-Webis} \cite{potthastStylometricInquiryHyperpartisan2018}, \textsc{Allcott \& Gentzkow} \cite{allcottSocialMediaFake2017}, \textsc{FakeNewsCorpus} \cite{pathakBREAKINGPresentingFake2019}) tend to be much smaller than those leveraging crowdsourcing. A common strategy to combat this, is to blend the `Article' and `Publisher' level labelling schemes (\textsc{FakeNewsNet/GossipCop} \cite{shuFakeNewsNetDataRepository2019}, \textsc{FakeNewsCorpus} \cite{pathakBREAKINGPresentingFake2019}, \textsc{MM-COVID} \cite{liMMCOVIDMultilingualMultimodal2020}). Either factual or misinformation articles are manually verified, and the complementary class is sampled from a set of publishers commonly associated with misinformation or factual articles, respectively.

A similar strategy is to blend the `Article' and `Claim' level labelling schemes. A claim made in an article is annotated for veracity in annotation, and its label is propagated to the entirety of the article (\textsc{FakeNewsNet/PolitiFact} \cite{shuFakeNewsNetDataRepository2019}, \textsc{CoAID} \cite{cuiCoAIDCOVID19Healthcare2020a}, \textsc{PolitiFact-Oslo} \cite{poldverePolitiFactOsloCorpusNew2023}).

The most consistent method for generating large, diverse corpora, however, proves to be using `Publisher-level labelling (\textsc{TSHP-17} \cite{rashkinTruthVaryingShades2017}, \textsc{Kaggle Fake News} \cite{risdalGettingRealFake2016}, \textsc{Some Like it Hoax} \cite{tacchiniItHoaxAutomated2017}, \textsc{Fake vs Satire} \cite{golbeckFakeNewsVs2018}, \textsc{QProp} \cite{barron-cedenoProppyOrganizingNews2019}), as discussed above.

Typically, the topics covered in the corpus are not further analyzed by dataset authors, although some datasets specifically focus on articles from various perspectives on the same events (\textsc{MediaEval15} \cite{boididouVerifyingMultimediaUse2015}, \textsc{PHEME} \cite{zubiagaExploitingContextRumour2017}, \textsc{BuzzFeed-Webis} \cite{potthastStylometricInquiryHyperpartisan2018}). In some cases, these instead influence the features used in automated misinformation classification (\textsc{Przybyła Credibility} \cite{przybylaCapturingStyleFake2020}).

Similarly, while most datasets are fairly general, some focus on specific domains. Very common are those focusing on social media or microblogging texts (\textsc{CREDBANK} \cite{mitraCREDBANKLargeScaleSocial2015}, \textsc{MediaEval15} \cite{boididouVerifyingMultimediaUse2015}, \textsc{Weibo15} \cite{maDetectingRumorsMicroblogs2016}, \textsc{Weibo17} \cite{jinMultimodalFusionRecurrent2017}, \textsc{WeChat} \cite{wangWeakSupervisionFake2020}). Another common domain involves celebrity rumors, typically annotated for verification rather than veracity, and also commonly sourced from social media posts (\textsc{Web Dataset Celebrity} \cite{perez-rosasAutomaticDetectionFake2018}, \textsc{FakeNewsNet/GossipCop} \cite{shuFakeNewsNetDataRepository2019}). During the COVID-19 pandemic, various health-related datasets were introduced (\textsc{FakeHealth} \cite{daiGingerCannotCure2020}, \textsc{MM-COVID} \cite{liMMCOVIDMultilingualMultimodal2020}, \textsc{FakeCovid} \cite{shahiFakeCovidMultilingualCrossdomain2020}, \textsc{CoAID} \cite{cuiCoAIDCOVID19Healthcare2020a}).

\subsection{Sources of Dataset Bias}

In this subsection, we discuss how specific operationalizations can introduce bias in the dataset, adversely affecting model generalization performance.

\subsubsection*{Differing Definitions}
Even among domain experts, there exists substantial disagreement on what does and does not constitute misinformation \cite{altaySurveyExpertViews2023}, with disagreements as to which degree content, medium or intent is relevant to defining misinformation \cite{gelfertFakeNewsDefinition2018a, yeeLimitsMachineLearning2025}. Recent systematic reviews have found that this disagreement has carried over to the computer sciences (see for example \citet{wuMisinformationSocialMedia2019, oshikawaSurveyNaturalLanguage2020, zhouSurveyFakeNews2020, aimeurFakeNewsDisinformation2023, bodaghiLiteratureReviewDetecting2024,xiaoChallengesMachineLearning2024}).
Indeed, the surveyed definitions of misinformation in Table \ref{tab:misinfo_datasets} seem to agree on basic properties of misinformation, but disagree on the specific forms. As a result, the forms of misinformation which are included can vary considerably. For example, misinformation forms like 'Satire' and 'Propaganda' are either explicitly included or excluded, proving to be especially divisive.

\subsubsection*{Inconsistent Label Sourcing} Another source of between-dataset variation, is the source of misinformation labels. While most datasets rely on domain experts, some use lay volunteers to verify content, either explicitly (\textsc{CREDBANK} \cite{mitraCREDBANKLargeScaleSocial2015}, \textsc{Weibo15} \cite{maDetectingRumorsMicroblogs2016}, \textsc{Weibo17} \cite{jinMultimodalFusionRecurrent2017}) or implicitly (\textsc{Some Like it Hoax} \cite{tacchiniItHoaxAutomated2017}).

Recently, datasets have started using many misinformation sources (\textsc{FakeCOVID} \cite{shahiFakeCovidMultilingualCrossdomain2020}, \textsc{MuMIN} \cite{nielsenMuMiNLargeScaleMultilingual2022}, \textsc{MCFEND} \cite{liMCFENDMultisourceBenchmark2024}). These can come from different countries and cultures, some of which are likely to disagree on their misinformation definitions. Furthermore, this requires aggregating the different misinformation labelling formats.

Most misinformation definitions require specific authorial intent to deceive. However, in some datasets this is missing  in the original content (\textsc{Lie Detector} \cite{mihalceaLieDetectorExplorations2009}, \textsc{FakeNewsAMT} \cite{perez-rosasAutomaticDetectionFake2018}), or ambiguous due to misinformation being defined as a lack of credible information (\textsc{FakeHealth} \cite{daiGingerCannotCure2020}, \textsc{PHEME} \cite{zubiagaExploitingContextRumour2017}, \textsc{FakeNewsNet/GossipCop} \cite{shuFakeNewsNetDataRepository2019}).

\subsubsection*{Few publishers}

Many datasets limit the number of publishers in either class. In some cases, this is due to deliberate scoping of the dataset (\textsc{BuzzFeed-Webis} \cite{potthastStylometricInquiryHyperpartisan2018}, \textsc{FakeNewsCorpus} \cite{pathakBREAKINGPresentingFake2019}), however in most cases this is due to publisher scarcity. Misinformation annotators, like Snopes, Politifact, GossipCop, etc., understandably tend to focus on verifiable misinformation pieces. As a result, datasets sampling annotations from these sources incur a large positive bias. A common strategy to counteract this is by including samples from a few mainstream publishers (\textsc{TSHP-17} \cite{rashkinTruthVaryingShades2017}, \textsc{MM-COVID} \cite{liMMCOVIDMultilingualMultimodal2020}, \textsc{CoAID} \cite{cuiCoAIDCOVID19Healthcare2020a}, \textsc{FakeNewsNet} \cite{shuFakeNewsNetDataRepository2019}).

An unwanted side effect of having a small, homogenous publisher set, is the introduction of a modelling shortcut; misinformation classifiers no longer need to analyze the veracity or intent of input content, but rather simply discriminate between a few publishers with unique idiosyncrasies. Similarly, in datasets where misinformation is constructed by editing factual information (\textsc{Lie Detector} \cite{mihalceaLieDetectorExplorations2009}, \textsc{FakeNewsAMT} \cite{perez-rosasAutomaticDetectionFake2018}), the labels can be inferred by discriminating between the stylistic preferences of the original texts' authors and those of the editors.

\subsubsection*{Few events or topics}

Similarly, many datasets sample content from a narrow time-span, or from a small set of events or topics. This can reduce the cost of generating labels, but will likely induce overfit in automated moderation systems trained on these corpora.

\subsubsection*{Focus on Obvious or Popular Misinformation}

In several of the discussed datasets, misinformation texts are collected based on user reports, or from third-party fact-checkers. These run the risk of introducing a selection bias, resulting in a dataset that is not representative of all produced misinformation.

A secondary effect of this, is that unverifiable content (e.g., those relying purely on opinion and speculation) are implicitly excluded. Some datasets explicitly exclude unverifiable content (\textsc{FakeNewsNet}  \cite{shuFakeNewsNetDataRepository2019}), whereas others include this as a specific category (\textsc{BuzzFeed-Webis} \cite{potthastStylometricInquiryHyperpartisan2018}). Most datasets do not discuss unverifiable cases, despite these forming a sizable part of produced online content (see Section \ref{sec:dataset_analysis_article_properties}).


\subsubsection*{Conclusion}

In short, we find that the realities of misinformation data collection results in many datasets making a trade-off between label quality and corpus size. As a result, these datasets introduce some bias, which we suggest as a primary reason for the reported brittleness of misinformation detectors under covariate shift. Given that these covariate shifts are practically guaranteed in online content or news, testing misinformation detectors before deployment for generalizability is crucial. Doing so, however, requires large, diverse datasets, which we have established is difficult to procure without bias. A related task, publisher reliability estimation, might provide an alternative.

\subsection{Publisher Reliability Datasets}

Related to the task of misinformation detection is publisher reliability estimation (see Section \ref{sec:related_works_publisher_reliability}). Given an article, or a set of articles, from some publisher, a reliability estimator has to predict the overall publisher reliability.

Publisher reliability is a broader concept than factuality, and considers many aspects of a publisher, which are not necessarily clear when analyzing articles or claims from a publisher in isolation. These aspects include framing, publisher political or editorial bias, intended audience, sourcing practices, funding, etc. All of these factors are analyzed on a large collection of a publishers works, and used to provide an indication of the trustworthiness of past and future releases. Ultimately, however, the factuality of produced articles is an important dimension of publisher reliability.

Much like the publisher-level misinformation labelling scheme discussed above, it does not preclude less reliable publishers producing reliable content, or \textit{vice versa}. It merely suggests that this is less likely to occur. Reliable publishers often produce sensationalist or subjective content to draw in readership, whereas unreliable publishers might intersperse their less reliable articles with more reliable ones to boost their perceived trustworthiness.

Implicitly, by using publisher-level labels as a proxy for article-level reliability, we (as well as many 'Publisher'-level datasets) make the assumption that the article-level factuality of an article from a reliable publisher is stochastically higher than that of an article from a less reliable publisher.


\subsubsection{Measuring Generalization with Publisher Reliability Labels}

Relative to misinformation datasets, for generalizability aspects, publisher-reliability datasets are far easier to produce at scale, and given publisher-level metadata, can be built specifically to enforce diversity in both publishers and text. Furthermore, articles can be collected across much longer time-spans, which naturally includes shifts in article topics or events.

Perhaps most importantly, however, if a large enough set of publishers is collected, the resulting dataset becomes a naturalistic view of published online content and (mis)information. Instead of a dataset including only verified misinformation, which are typically the least ambiguous or popular cases due to the selection bias of third-party fact-checkers, the dataset is more aligned with online content as it would appear post deployment (Section \ref{sec:dataset_analysis}). As a result, statistics about model evaluation are more representative, and model developers can derive stronger conclusions.


In this article, we propose using a publisher-level reliability estimation dataset for the purpose of evaluating the generalizability of article-level misinformation detectors. While this runs the risk of tarring all articles from a publisher with the same brush, we believe the size and diversity of the dataset, along with access to publisher-level metadata, can offset the induced bias and still allow for conclusive inferences about model behavior under distribution shift.

Specifically, we assume that the effect of covariate distributional shifts on the predictive quality of a model is positively correlated between the two labeling approaches. In other words, we assume that model performance degrades under the same distributional shift in both labeling set-ups. Thus, implicitly, we assume that the level of robustness to distributional shifts on a dataset like \texttt{misinfo-general} serves as a good indicator for robustness in article-level misinformation detection.



\section{The \texttt{misinfo-general} Dataset}
\label{sec:misinfo-general}

Here we introduce \texttt{misinfo-general}, a benchmark for testing the generalization capacity of misinformation detection models, built on top of a series of noisy publisher-level datasets. While best suited for publisher reliability estimation models, we instead use the publisher labels as a proxy for article labels.

Based on the prior discussion, we foresee two sources of bias, (1) labels might not be accurate at the article level, and (2) models will learn to infer the article's publisher and its label instead of inferring the label from the article. We take the following steps to mitigate these biases as much as possible:
\begin{enumerate}
  \item relabelling existing articles (Section \ref{sec:misinfo-general_labelling})
  \item masking or removing publisher identifiable text in articles (Section \ref{sec:misinfo-general_processing})
  \item removing any article- and sentence-level duplicates (Section \ref{sec:misinfo-general_processing})
  \item masking self-references, along with other PII (Section \ref{sec:misinfo-general_processing})
\end{enumerate}

In Section \ref{sec:dataset_analysis_publisher_prediction} we show that these pre-processing steps have made publisher identification from articles alone difficult.

In this section, we describe how we gather the dataset content and labels, and generate any additional metadata. Later sections make use of article-level metadata, and we specifically test for model overfit to publisher style (Section \ref{sec:Generalization-forms} \& Section \ref{sec:results}), we show including a diverse set of publishers is beneficial to generalization performance (Section \ref{sec:analysis_dataset_effect}), and we try to find publishers with high degrees of mislabelling by assessing the necessity of model memorization (Section \ref{sec:dataset_analysis_mislabelling}).

\subsection{Article Provenance}
\label{sec:misinfo-general_article_provenance}

All raw articles come from the various \textbf{Ne}ws \textbf{La}ndscape (NELA) corpora produced by the \href{https://melalab.github.io}{MELA} lab\footnote{\url{https://melalab.github.io}} \cite{horneSamplingNewsProducers2018, norregaardNELAGT2018LargeMultiLabelled2019a, gruppiNELAGT2019LargeMultiLabelled2020, gruppiNELAGT2020LargeMultiLabelled2021,
gruppiNELAGT2021LargeMultiLabelled2022, gruppiNELAGT2022LargeMultiLabelled2023}. The corpora cover 2017--2022 (6 iterations) almost continuously, with articles from a diverse group of publishers. In their original form, the 6 iterations together consist of $7.2$ million long-form articles.

The original authors' goal was to study the dynamic behavior of news and news publishers. They deemed existing corpora inadequate for their goals, because of (1) a small, relatively homogenous collection of articles or publishers, (2) too narrow a focus on specific events, (3) bias towards popular publishers, and (4) limited ground truth labelling \cite{horneSamplingNewsProducers2018, norregaardNELAGT2018LargeMultiLabelled2019a}.

\subsection{Publisher Labelling}
\label{sec:misinfo-general_labelling}

From the 2018 iteration onwards, the NELA datasets come with publisher-level labels. However, due to inconsistencies across dataset iterations and the frequency of labelling errors, we chose to relabel the dataset completely.

Similar to the initial NELA corpora labels, we scraped \href{https://mediabiasfactcheck.com/}{Media Bias/Fact Check}\footnote{\url{https://mediabiasfactcheck.com/}} (MBFC). MBFC is a curated database of news publishers, with thorough analyses of publisher origins, bias, and credibility. Despite being run by lay volunteers, MBFC labels correlate well with professional fact-checking sources \cite{kieselSemEval2019Task42019, broniatowskiTwitterFacebookPosts2022, pratelliStructuredAnalysisJournalistic2022}. MBFC labels have been used in many earlier works \cite{rashkinTruthVaryingShades2017,balyPredictingFactualityReporting2018,balyWeCanDetect2020,burdissoReliabilityEstimationNews2024,casavantesPropitterTwitterCorpus2024,szwochLimitationsLargeLanguage2024}. We use the metadata available as of Oct. 2024, well after the final publication dates of articles in the corpus.

Using the URL domain of the scraped articles, We first mapped all articles to a consistent set of publishers before removing any publishers known to be news aggregators or social media sites. This gives an article-publisher mapping that is consistent across dataset iterations, and removes cases of where articles were republished on different sites. Each publisher was linked to a publisher in the scraped MBFC database. We provide further detail in Appendix \ref{app:dataset_relabelling}. Ultimately, we identified $488$ distinct publishers, many of which were falsely attributed in NELA's original set of publishers. The metadata available for each publisher is provided in Appendix \ref{app:publisher_metadata}.

The MBFC database is dynamic, and it does happen that the publisher label or metadata annotations change\footnote{See \url{https://mediabiasfactcheck.com/changes-corrections/}}. Usually, this presents as a relatively minor change in political bias. During the data collection and processing period (Jan. 2017-Oct. 2024), we found 20 instances where the change was substantive (see Appendix \ref{app:metadata} Table \ref{tab:publisher_label_changes}). In the majority of cases (12/20), this resulted in a previously \reliable publisher failing too many fact checks, resulting in their rating being downgraded to \unreliable. Ultimately, all these cases are due to additional information about the publishers' editorial practices coming to light, rather than those practices changing. In 5 cases publishers either corrected articles with failed fact checks or shifted their editorial policies, resulting in a label shift from \unreliable to \reliable.

\subsection{Data Processing}
\label{sec:misinfo-general_processing}

Beyond errors in the article-publisher and publisher-label mappings, the texts themselves frequently contain duplicates or scraping errors. Of the $6.7$M re-labelled articles, roughly $\approx22\%$ or $1.5$M articles were duplicates. Many of the remaining unique articles were deemed malformed or semantically void. These contain either very little text, substantial amounts of markup or include too many special tokens to be human-readable. We filter these using a few simple rules (see Appendices \ref{app:dataset_deduplication} and \ref{app:cleaning_filtering}). Altogether, we remove approximately $\approx43\%$ of all downloaded articles. The final dataset contains $4.2$ million cleaned articles.

In the remaining texts, we mask various forms of private or identifiable information (PII), both to enhance safety and reduce the number of available classification `shortcuts'. We furthermore standardize the copyright masking procedure introduced in \citet{gruppiNELAGT2019LargeMultiLabelled2020}. This introduces $4$ new special tokens: \texttt{<copyright>} replacing NELA's repeated \texttt{@} tokens, \texttt{<twitter>}, \texttt{<url>} and \texttt{<selfref>} for any self-references.

Despite our efforts, the datasets retain a level of `noise' customary to data sourced from the internet. For example, articles from the same publisher tend to contain unique by-lines, attribution messages, or donation requests. Further cleaning efforts might reduce the realism of the benchmark.

\subsection{Topic Clustering}
\label{sec:misinfo-general_topic_clustering}

One of our aims is to test model generalization across different events and topics. To discover these, we used a modified variant of \texttt{BERTopic} \cite{grootendorst2022bertopic} with a \texttt{gte-large}\footnote{\href{https://huggingface.co/thenlper/gte-large}{https://huggingface.co/thenlper/gte-large}} \cite{liGeneralTextEmbeddings2023} backbone. This produced thousands of event clusters for every dataset iteration, each with a TF-IDF representation vector. We aggregate these events into overarching topics by applying spectral clustering to the adjacency matrix induced by the inter-event cosine similarity of the TF-IDF matrix. We arbitrarily limit the number of topics to 10, each with varying numbers of events in them. This process is further described in Appendix \ref{app:topic_clustering}.

This largely mimics the process used in \citet{przybylaCapturingStyleFake2020}, and extends the work of \citet{littererWhenItRains2023} on identifying `news storms' in the NELA corpora to a larger time-span, and a larger set of publishers.

\section{Generalization Taxonomy}
\label{sec:Generalization-forms}

In this section, we describe various dimensions along which we believe covariate shifts likely to occur, and which are feasible to simulate using \texttt{misinfo-general}. We consider a total of $6$ specific generalization axes.

\textbf{Time} based generalization measures the extent to which changes in publisher style affect a model's predictions. The publishers considered in each split should be held constant to avoid confounding with different publishers.

Evolution of article content will also impact performance. We focus on two specific forms of such change: (1) due to spontaneous \textbf{events}, which we define as news-worthy happenings with a definite and narrow time-span, or (2) due to evolving \textbf{topics}, which we define as large, overarching collections of events that remain relatively static over a long period. Across these events and topics, we expect markedly different language.

The distribution of publishers is also expected to change between training and inference time. All \textbf{publishers} exhibit some form of editorial bias or style, which can be memorized by classification models. While models should use style to inform moderation decisions, they should also not overfit to stylistic idiosyncrasies. One related, usually implicit, expectation of misinformation detectors is a robustness to different \textbf{political biases} or \textbf{misinformation types}. Predictions ought to be based on a publisher's intent, not their norms and values. By excluding these from training, we can test a model's ability to generalize to different classes of publishers.

\subsection{Data Splits}

To operationalize these generalization axes, we build 6 (+1 baseline) train/test splits of the dataset using the publisher-level metadata available to us. Each split is meant to simulate one of the above described covariate shift scenarios, while ensuring minimal cross-scenario confounding.

Throughout, we approximate the same $70$/$10$/$20$\% article proportions per training/validation/test split, respectively. The validation split, used for early stopping, is sampled i.i.d. from the training set. For all scenarios, we repeat each split independently for each dataset year, for a total of 6 times. The only exception is the `Event' axis, for which we combine all years into a single dataset.

Briefly, we construct splits (schematically displayed in Table \ref{tab:generalisation_forms}) for the scenarios as follows:

\begin{table}
  \centering
  \caption{A schematic overview of the generalization taxonomy. The left columns provide relevant generalization category and axis, whereas the right columns provide examples of in domain and out-of-distribution article sets.}
  \label{tab:generalization_taxonomy}
  \input{tables/generalisation_taxonomy}
\end{table}

{
\begin{enumerate}
  \setcounter{enumi}{-1}
  \item \textbf{Uniform}: standard stratified random splitting of articles into disjoint article sets. No article meta-data is used
  \item \textbf{Time}: the training set consists of a single dataset year, while the test set contains articles from publishers seen during training in all other dataset years. This tests within publisher variation
  \item \textbf{Event}: the dataset has been annotated for several thousands of events, but we focus on a singular one: the COVID-19 pandemic. We reserve all articles containing any related keywords for testing, and we train on all non-COVID articles
  \item \textbf{Topic}: we reserve the $k$ smallest topic clusters for the test set, such that these contain roughly $~20\%$ of all articles, and we train on the remaining articles
  \item \textbf{Publisher}: similarly, we reserve the $k$ least frequent publishers for the test set, such that these contain roughly $~20\%$ of all articles, and we train on the remaining articles
  \item \textbf{Political Bias}: we reserve all articles from either all `Left'- or `Right'-biased publishers for testing, and train on articles from the opposite political bias, along with any `Center'-biased publishers
  \item \textbf{Misinformation Type}: similarly, we reserve all articles from either all `Questionable Source' or `Conspiracy-Pseudoscience' publishers for testing, and train on articles from the other misinformation class. We use an i.i.d. split of reliable articles to ensure a similar class distribution in all splits
\end{enumerate}
}

We include a substantially expanded description of each split's construction in Appendix \ref{app:generalisation_forms}. In the `Topic' and `Publisher' splits were constructed by sampling from the smallest topics and publishers. This was to ensure that all splits have approximately the same size while simultaneously maximizing the diversity of the held-out test sets. This could introduce a bias towards the more prolific publishers and topics; however, (1) this bias is already present in the training data (see Appendix \ref{app:determinants}, Tables \ref{tab:determinants_coefficients} and \ref{tab:determinants_coefficients_topic}, parameter `train count'); and (2) we do not believe this had an undue amount of influence on the quality of the training models (see Section \ref{sec:analysis_dataset_effect}).

It is important to note that from the model's perspective, each scenario seems identical. The same labels are present in each split, with roughly the same article counts in the same class proportions. Without additional context, one should expect similar performance across these splits.


\section{Experiments and Results}
\label{sec:results}

To showcase the utility of \texttt{misinfo-general} for model training and evaluation, we use a simple yet powerful baseline model. Specifically, we fine-tune an instance of \texttt{DeBERTa-v3}\footnote{\href{https://huggingface.co/microsoft/deberta-v3-base}{https://huggingface.co/microsoft/deberta-v3-base}} \cite{heDeBERTaV3ImprovingDeBERTa2022} where we reset the pooler and classification weights but freeze the model's remaining weights. To enable using dataset-specific tokens, we allow the token embedding layer to train with a very low learning rate. The model's pre-training data included a closed-source news dataset (CC-News), dated between September 2016 and February 2019 \cite{liuRoBERTaRobustlyOptimized2020}, and thus should be easily adapted to \texttt{misinfo-general}. Similar architectures have shown surprisingly adequate performance on other benchmark datasets, including various NELA versions \cite{pelrineSurprisingPerformanceSimple2021,zhouHiddenBiasesUnreliable2021a, razaFakeNewsDetection2022}.

We fine-tune the models on the different splits outlined in Section \ref{sec:Generalization-forms}. We keep the hyperparameters and compute budget constant (which were tuned on the validation sets of the `Uniform' splits), which we outline in Appendix \ref{app:training_details}. Training occurs at the article level, using publisher-level labels. We binarize the article publisher's MBFC label for training labels: all `Questionable Source', `Conspiracy-Pseudoscience', and `Satire' publishers were deemed \unreliable, and all others \reliable. Other publisher-label mappings have been used in other works, and is deserving of future research for this dataset.

To assess model performance at the article level, we employ the F1-score computed independently for each class along, with the Matthews Correlation Coefficient (MCC). The F1-score's interpretation is largely dependent on the class proportion \cite{flachPrecisionRecallGainCurvesPR2015}, making it less suited to comparison across experiments, whereas MCC is more robust to this \cite{chiccoAdvantagesMatthewsCorrelation2020, chiccoInvitationGreaterUse2022}. MCC is 0 for random performance, and 1 only for perfect classification.

\begin{table}[t]
  \caption{Article-level classification performance comparing performance on the ID and OoD evaluation sets. The top row uses uniform splitting for both (OoD = ID), serving as a baseline value. `Time' based splitting has strongly varying class proportions, making F1 values inappropriate.}
  \label{tab:generalisation_forms}
  \input{tables/generalisation_forms}
\end{table}

\subsection{OoD Generalization}
\label{sec:results_ood_Generalization}

Table \ref{tab:generalisation_forms} displays the article level classification results for the various generalization splits outlined in Section \ref{sec:Generalization-forms}. The larger the deviation between the in distribution (ID) articles in the validation set and the out-of-distribution (OoD) articles in the test set, the worse we consider the model's generalization performance.

Firstly, we note that classification performance falls short of desired. While the F1-score for the \reliable class tends to be high (in the range of $\mathbf{0.85-0.95}$ at a $\sim60$\% class proportion), classifying \unreliable articles is considerably more difficult---a trend that holds consistently across generalization forms. This is largely due to low recall scores for the \unreliable class. This is especially surprising given the high accuracy scores reported for similar models on other misinformation datasets.

We see no degradation in performance when applying the model to articles from an unseen event (here, the COVID-19 pandemic). Despite the introduction of many unseen terms to the articles' vocabulary, it appears the manner in which established publishers discuss this new event deviates little from preceding articles.

Both `Publisher' and `Topic' splitting show moderate decreases in MCC scores, carried primarily by a decrease in the F1-scores for the \unreliable class. Generalization to completely unseen publishers or topics, cases where one would expect distinctly different linguistic style or vocabulary, is more challenging. The magnitude of this performance degradation, however, is smaller than we initially expected. We attribute this to two effects:

{
\begin{enumerate}
  \item More mainstream, prolific publishers are obscuring performance on publishers with fewer articles (see Appendix \ref{app:publisher_statistics}). We correct for this effect by including a publisher-level analysis in Section \ref{sec:analysis_determinants} \\
  \item The training data is heterogeneous enough for the models to learn generalization across publishers. We test for this in Section \ref{sec:analysis_dataset_effect}
\end{enumerate}
}
Since it is conceivable that different publishers prefer particular topics, we compute a correlation between the produced test sets. While we find a small but consistent overlap between the `Publisher' and `Topic' test sets, we do not believe this alone accounts for the similarity in performance (see Appendix \ref{app:test_test_correlations}).

The final two generalization axes exclude a particular misinformation type or political bias from the training set. For the former, we can see little to no effect when removing the `Conspiracy-Pseudoscience' class of articles, but a drastic one if removing `Questionable Source' articles. We posit this is due to the `Questionable Source' being the class of articles written with the explicit purpose of mimicking \reliable publishers, whereas `Conspiracy-Pseudoscience' tends to discuss completely separate topics. In other words, the conspiracy or pseudo-scientific articles tend to be easier to identify as \unreliable.

For the `Political Bias' generalization axis, we see an inability to generalize to opposing political biases. Training on center and right biased articles sees a $\mathbf{0.19}$ drop in MCC, whereas training on center and left yields a drastic $\mathbf{0.37}$ drop. While this is a form of publisher splitting, in both cases the magnitude of the degradation is substantially larger. Especially for transfer to right-biased articles, there exists a drop for both \reliable and \unreliable classification, indicating that it is more challenging for the model to determine article reliability.

\subsection{Generalization across time}
\label{sec:performance_across_time}

When applying the models to unseen years, we find the models to be surprisingly robust, as shown in Table \ref{tab:uniform_across_years}. At the article level, despite consistent degradation in performance, proximal years tend to achieve similar scores. Only in very distant years does performance degrade dramatically.

We speculate that these differences are due to differences in the various dataset iterations, while publisher style or idiosyncrasies being relatively static. For example, all models not trained on the 2017 iteration perform poorly on the 2017 iteration (between $\mathbf{0.26}$ and $\mathbf{0.34}$ MCC), whereas the 2020–2022 editions perform reasonably well on each other's years. Indeed, visually, Table \ref{tab:uniform_across_years} correlates strongly with Appendix \ref{app:publisher_statistics} Table \ref{tab:publisher_overlap}, showing the amount of overlap in publishers across dataset years.

\begin{wraptable}[15]{r}{0.45\textwidth}
  \centering
  \vspace{-1.4em}
  \caption{Article level MCC scores of models trained with uniform splitting on different years of the dataset.}
  \label{tab:uniform_across_years}
  \input{tables/uniform_across_years}
\end{wraptable}

\subsection{LLM Performance}

We compare the performance of the fine-tuned models to that of \texttt{llama-3-8b-instruct}\footnote{\href{https://huggingface.co/meta-llama/Meta-Llama-3-8B-Instruct}{https://huggingface.co/meta-llama/Meta-Llama-3-8B-Instruct}} \cite{dubeyLlamaHerdModels2024}, prompted to determine reliability of an article in a 0-shot setting with $512$ token context (see Appendix \ref{app:llm-inference}).

Despite the LLMs parameter count and the recency of its pre-training data, we find \texttt{llama-3-8b-instruct} to be inferior to the fine-tuned models as for the purpose of article-level reliability classification. It manages an MCC of $\mathbf{0.25}$, compared to our fine-tuning models achieving $\mathbf{0.46}$ on ID years, and $\mathbf{0.33}$ on OoD years.

The use of this modestly sized LLM already incurs a computational cost far greater than that of the fine-tuning models. In our experiments, using a single A100 GPU, the LLM took \textasciitilde$70$ hours to yield a prediction for all articles in the corpus, whereas each of the fine-tuning models were trained and evaluated in \textasciitilde$12.5$ hours. For deployment scenarios where the amount of compute necessary for inference far exceeds that of training, this difference will likely be more pronounced, and platforms with a large influx of text (i.e., social media networks) will need to balance the substantial computational overhead of model inference with classification performance.

\begin{wraptable}[16]{r}{0.45\textwidth}
  \centering
  \caption{Performance of `Uniform' fine-tuned decoder-only models, compared to several SoTA reasoning LLMs (via API), on a small, stratified subset of the dataset. The left column provides the name of the model, and the rows below the thinking budget provided to the model.}
  \label{tab:reasoning_models}
  \input{tables/reasoning_models}
\end{wraptable}

While it is conceivable that larger, more recent language models might achieve strong misinformation detection performance, due to the size of the corpus and length of the articles, we did not experiment further with such models. Additionally, recent experiments with ICL misinformation detection have resulted in subpar performance \cite{yangAccuracyPoliticalBias2025}.

\subsubsection{Reasoning LLMs}
\label{sec:llm-performance-reasoning-llms}

We additionally experiment with some reasoning models \cite{deepseek-aiDeepSeekR1IncentivizingReasoning2025custom, comaniciGemini25Pushing2025custom}. Unlike standard LLMs, these models are post-trained for reasoning tasks, and produce long `thoughts' before answering a question.

We compare these models against the fine-tuned models on a small, stratified subset of the entire dataset\footnote{Due to budget constraints} that contains two articles per publisher-topic combination for a maximum of 120 articles per publisher, totalling 28k articles. Similar to the above LLM and fine-tuning experiments, we only provided 512 tokens of context.

Table \ref{tab:reasoning_models} shows the MCC and F1 scores these models achieve. We find that these models can achieve significantly better performance on this publisher-topic stratified subset, primarily through higher \unreliable precision scores.

Again, it should be noted that the reasoning models have orders of magnitude more parameters and pre- and post-training data. As a result, it is plausible that these models have some knowledge about the articles and the events they depict, and are capable of placing individual articles in substantially more context than the fine-tuned models can. This becomes readily apparent when analyzing the reasoning models' `thoughts'. These contain frequent references to quoted publishers and entities, whose reliability is known \textit{a priori}, and the models seem to have a keen understanding of how these interact with reliable journalistic practices.

As a result, the results are likely not directly comparable to the purely inductive, fine-tuned models, with the reasoning models being able to apply a mixture of generalizable rules and external knowledge. This likely means that their generalization performance is overestimated, and one might expect the same set of issues identified in Sections \ref{sec:introduction} \& \ref{sec:related_works} to occur: the models are not being evaluated for their performance on OoD data.

\section{Analysis of Model Generalization}
\label{sec:analysis}

\subsection{Determinants of Performance}
\label{sec:analysis_determinants}

The analysis of our results, thus far, has been constrained to article-level classification. While this reflects how misinformation classifiers interact with articles, it does not match how we annotate the dataset and can obscure performance on smaller, less mainstream publishers. Ideally, as highlighted by \citet{balyIntegratingStanceDetection2018} and \citet{burdissoReliabilityEstimationNews2024}, classification performance is also evaluated at the publisher-level, testing which publisher properties aid or interfere with misinformation detection.

\begin{wrapfigure}[23]{r}{0.45\textwidth}
  \centering
  \includegraphics[width=0.45\textwidth]{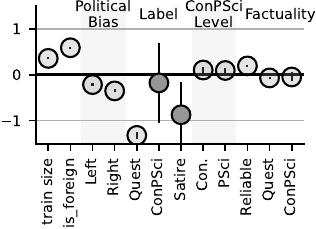}
  \caption{Coefficients of the determinants model, expressed as effect sizes. Circles are centered about the effect size, with lines giving the 95\% confidence interval.}
  \label{fig:determinants}
\end{wrapfigure}

To that end, we employ a binomial logistic regression on the average publisher-level accuracy\footnote{Defined as the ratio of correctly predicted articles to all articles for a publisher, i.e., true positive rate, recall.} to assess which aspects of a publisher determine the achieved accuracy score (for details and full model specification, see Appendix \ref{app:determinants}). Unlike standard logistic regression, the dependent variable is modelled as a ratio.

\begin{wraptable}{r}{0.45\textwidth}
  \vspace{-10.0em}
  \centering
  \caption{Median predicted publisher-level accuracies averaged over combinations of the MBFC label (rows) and political bias (columns). The row headers correspond to (R) reliable, (Q) Questionable Source, (C) Conspiracy Pseudoscience, and (S) Satire.}
  \label{tab:emms_uniform}
  \input{tables/uniform_predictions_emms}
\end{wraptable}

In Figure \ref{fig:determinants} we show coefficient magnitudes for several important determinants, expressed as effect sizes \cite{chinnSimpleMethodConverting2000,lampinenCanLanguageModels2022}. A positive effect size indicates that the variable increases the odds of accurate classification, \textit{ceteris paribus}.

The size of the training set has a large positive effect, with a $\mathbf{1.91}$ multiplicative increase in the odds for each 10-fold increase in training samples. Thus, a publisher with 1000 articles in the training set is $\mathbf{3.82}$ times more likely to have correct classification in the test set than a publisher with only 10 articles. Foreign publishers also prove easier to identify.

Relative to center-biased publishers, publishers on the left or right sides of the political spectrum see slightly degraded performance, even after controlling for the publisher label. Moreover, classification on the \unreliable classes suffers more than on \reliable publishers. Since the odds ratios or effect sizes are difficult to interpret, we also provide the estimated marginal mean publisher-level accuracy for those combinations in Table \ref{tab:emms_uniform}.

Despite right biased \unreliable sources being far more prevalent in the training data, for both the `Questionable Source' and `Conspiracy-Pseudoscience' classes, model performance is noticeably worse than on left and center biased sources. This somewhat confirms the results in the `Political Bias' rows of Table \ref{tab:generalisation_forms}: models struggle disproportionately to discriminate between reliable and \unreliable right-biased articles.

We see a slight positive correlation with the MBFC `Conspiracy-Pseudoscience' level. The higher the value, the further the publishers' articles tend to deviate from convention. As a result, strongly conspirational or pseudo-scientific publishers are $\mathbf{4.84}$ and $\mathbf{4.72}$ times more easily identified than publishers where this effect is weaker.

Finally, when looking at the factuality level (the propensity for a publisher to publish factual articles) we find a positive interaction for reliable articles, and weakly negative interactions for unreliable articles. The more a publisher goes against the expectation (factual for reliable for publishers, false for unreliable), the more difficult it becomes to disambiguate the source.

\subsection{Effect of Publisher Diversity}
\label{sec:analysis_dataset_effect}

\begin{wrapfigure}[21]{r}{0.45\textwidth}
  \centering
  \vspace{-1.0em}
  \includegraphics[width=0.45\textwidth]{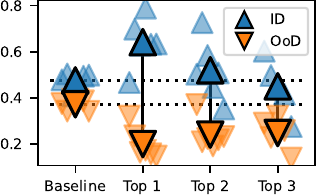}
  \caption{MCC scores for different `Publisher' split test sets with only the top-1, 2 or 3 most prolific publishers retained per publisher class. The `Baseline' column corresponds to the standard publisher splitting used in described in \ref{sec:Generalization-forms}. The lightly shaded shapes provide a value for each year of the dataset, the solid shapes their average. The blue upward triangles are ID publishers, the orange downward triangles instead represent OoD publishers.}
  \label{fig:dataset_effect}
\end{wrapfigure}

Here we test to what extent the diversity of the publishers present in the training data has an effect on the generalization capacity of the models. We re-run the `Publisher' split experiment with smaller training sets, constrained to only the most prolific publishers. Specifically, for each MBFC label, we only include the top-n most frequent publishers in the training data, while leaving the test set untouched. While this reduces the amount of variation in publishers considerably, it minimally affects the total amount of data present. Each training set still consists of hundreds of thousands of articles.

Figure \ref{fig:dataset_effect} displays the generalization gap (in terms of MCC) induced when increasing publisher homogeneity. Especially when limiting performance to the Top-1 most common publishers, the models show increased overfit to the training set. Where the `Publisher' split saw a $\mathbf{0.1}$ MCC delta, this increased to an average degradation of $\mathbf{0.5}$. As the number of included publishers increases, the generalization gap decreases and starts to converge to the previously seen `Publisher' values. Notably, the variance in values is also substantially higher in the limited publisher settings.

From this, we conclude that (1) the splitting described in Section \ref{sec:Generalization-forms} to have minimally altered the heterogeneity present in the dataset, and (2) the models improve with publisher heterogeneity. The former finding suggests that the underestimation of the generalization gap will be especially egregious in datasets with a small pool of publishers (e.g., those that sample from a single reliable source to boost label balance). The latter, instead, provides some initial evidence for the utility of using large, diverse publisher-level datasets for pre-training article-level misinformation detectors; while fine-tuning on high-fidelity labels is likely necessary, using distantly supervised datasets might encourage more robust models before fine-tuning.

\section{Analysis of Publisher-Level Labelling}
\label{sec:dataset_analysis}

\subsection{Publisher Identifiability}
\label{sec:dataset_analysis_publisher_prediction}

The use of publisher-level labels as a form of weak supervision, especially in misinformation detection, can lead to models overfitting to publisher styles instead of article veracity. This was shown to be a serious concern by \citet{zhouHiddenBiasesUnreliable2021a}, and efforts to mitigate this effect at the data level were discussed in Section \ref{sec:misinfo-general}. Despite this, in Sections \ref{sec:results_ood_Generalization} and \ref{sec:analysis_determinants} we still found models to overfit to specific publishers and publisher classes, and in Section \ref{sec:analysis_dataset_effect} we found a negative correlation between publisher diversity and the magnitude of the generalization gap.

As such, here we directly test the identifiability of the publisher from article content by replacing the misinformation labels with a unique identifier for each publisher. In other words, instead of classifying into the set $\{\text{\reliable}, \text{\unreliable}\}$, the model classifies into the set of all possible publishers.

Using the same learning algorithm, we find this to be a substantially more difficult task. While models exhibit above random article-level performance, with an average MCC score of $\mathbf{0.18}$, this is much lower than scores achieved with misinformation labels. Furthermore, when aggregating F1-scores across classes proportionally according to publisher frequency (micro) we get $\mathbf{0.14}$,  whereas with a flat average (macro) we obtain a mere $\mathbf{0.04}$ F1. In short, while it is possible to predict the publisher from an article with above random performance, this is only really possible for the most prolific publishers, and this cannot entirely explain performance in misinformation classification.


\subsection{Publisher Memorization}
\label{sec:dataset_analysis_mislabelling}

In this subsection, we analyze to which extent models need to memorize specific publishers. If there exists a lot of disagreement between the label of an article and the label assigned to its publisher, it is likely impossible to generalize to the unseen publisher from seen publishers; the labels of similar publishers clash. In this case, for classification to be successful, it is necessary for the misinformation detector to memorize publisher idiosyncrasies.

\begin{figure}
  \centering
  \includegraphics[width=1.0\textwidth]{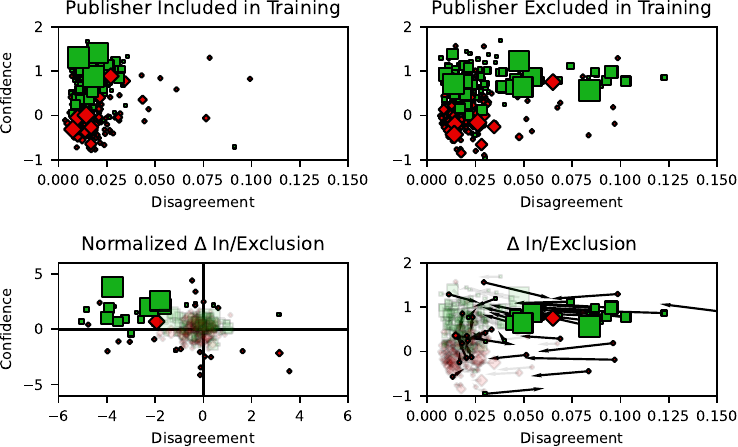}
  \caption{The average article-level confidence and disagreement for different publishers. The top left panel shows scores when publishers are included in training, whereas the top right panel shows scores when publishers were excluded. The bottom left panel shows their difference, normalized. Only publishers with a shift magnitude above $2$ standard deviations (i.e., significantly far from origin) are shown with low opacity. The bottom right panel shows the direction of differences for significantly shifted publishers. In all panels, the size of the circles are proportional to a publisher's size in the dataset. Green colored squares represent \reliable publishers, whereas red colored diamonds represent \unreliable ones.}
  \label{fig:dataset_map}
\end{figure}

Inspired by the works of \citet{pleissIdentifyingMislabeledData2020}, \citet{jenkinsOrderedSampleConsensus2023} in automated mislabelling detection, and \citet{swayamdiptaDatasetCartographyMapping2020} on diagnosing dataset issues using `dataset cartography', to estimate the necessity of memorization of specific publishers, we run an experiment comparing the average article confidence (mean logit assigned to the correct class) and disagreement (variance of logit assigned to the correct class) of publishers when included or excluded from the dataset. If there is a large shift in either the confidence or disagreement of a publisher when in- or excluded, this might indicate the publishers' articles' labels are not aligned with those of similar publishers.

We rerun the 2021 uniform split experiment 15 times. We exclude each publisher in 5 runs, at random, while taking care to minimize the number of exclusion set co-occurences and stratifying the exclusion across fine-grained MBFC publisher classes. As such, there should always be similar publishers available to excluded ones.

Figure \ref{fig:dataset_map} shows the effect of exclusion on the average confidence and disagreement scores for all publishers. Overall, and unsurprisingly, including a publisher during training increases average article confidence and decreases disagreement. However, for most publishers, the shift between training in- or exclusion is small, and likely attributable to the inherent stochasticity of mini-batch training. We assume that these publishers' labels can largely be learned from similar publishers, and that these align well with each other.

While there are significant shifts, these mostly occur for the largest, typically well-performing \reliable, publishers, and take the form of significantly increased disagreement and decreased confidence when excluded. These include sources which MBFC believes to produce typically reliable, but sensationalist, subjective news (e.g., The Sun, The Daily Mirror), or anti-US propaganda sources (e.g., Pravda Report, Asia Pacific Research), as well as highly reputable sources (e.g., CBS News, BBC)\footnote{These categorizations originate from MBFC, and do not reflect the authors' opinions.}. Comparatively, most large \unreliable publishers see far smaller shifts.

There are substantially fewer significant shifts in the other quadrants (Figure \ref{fig:dataset_map}, bottom left panel), and those that do show up tend to be for much smaller publishers. Publishers that see a significant \textit{reduction} in confidence when included in training do exist, although these comprise a small minority with typically very few articles. Looking more closely at such publishers, these include cases where site ownership changed during article collection (Viral News Network, Infinite Unknown), or whose articles are extremely noisy (Alternative Media TV), which would serve as good candidates for removal. We also find particularly difficult cases here, like neutral, objectively written articles promoting climate change denial (Climate Etc)\footnote{Idem.}.

All in all, while there appears to be some ambiguity in article labels, largely due to publisher editorial biases, we find no evidence of mislabelling beyond a level expected for a corpus scraped from the internet. Despite having missed some publishers in the data cleaning phase (see Appendix \ref{app:dataset_processing}), these represent a small minority of all publishers, and collectively contain a small minority of all included articles.

\subsection{Article Properties}
\label{sec:dataset_analysis_article_properties}

In this subsection, we automatically analyze various properties of our dataset at the article level, both to assess their presence for different classes of publishers, and their correlation with news reliability labels.

\begin{wrapfigure}{r}{0.45\textwidth}
  \centering
  \vspace{-1.0em}
  \includegraphics[width=0.45\textwidth]{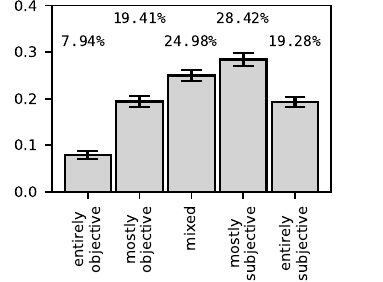}
  \caption{The estimated proportions for each subjectivity level in \texttt{misinfo-general}. Errors bars give the $95\%$ Agresti-Coull binomial proportion confidence interval \cite{agrestiApproximateBetterExact1998}}\label{fig:dataset_analysis_subjectivity_proportions}
\end{wrapfigure}

\paragraph{Subjectivity analysis} The first property we annotate for is subjectivity. Objective news presents facts in a neutral, unbiased manner, and is commonly considered the antithesis of hyperpartisan or misinformation news, which is written specifically to incite an emotional response from readers, thereby inducing sharing \cite{bojicMaintainingJournalisticIntegrity2024}. Subjectivity has shown some promise as a feature in discriminating reliable and unreliable news \cite{jeronimoFakeNewsClassification2019}. Despite this, reliable publishers also produce subjective text, likely to drive engagement. This can make articles unverifiable, hampering article-level labelling.

To assess the degree of objectivity, we ask \texttt{ChatGPT-4o-mini} to provide a rating for an article using a $5$-point Likert scale, ranging from entirely objective to entirely subjective. While by no means SoTA, similar setups have shown reasonable performance in prior work \cite{galassiNotebookCheckThatLab2023, strussNotebookCheckThatLab2024, shokriSubjectivityDetectionEnglish2024}.

Figure \ref{fig:dataset_analysis_subjectivity_proportions} shows the estimated proportions of each subjectivity level. Despite reliable news being in the majority, most ($\sim\mathbf{73}\%$) articles have a subjectivity level of `Mixed' or higher. In fact, `Mostly Subjective' articles seem to be most common.

Upon inspection of the dataset, these annotations seem to match our findings. While unreliable news is substantially less likely to present itself as objective, reliable publishers still publish a plethora of discussion and opinion pieces. This is especially true for publishers with a more pronounced political bias.

We repeat the binomial logistic regression analysis used in Section \ref{sec:analysis_determinants} to determine what publisher-level properties correlate with subjectivity. Unlike earlier, where the ease of classification correlated strongly with the form of misinformation, article subjectivity tends to correlate strongly with publisher political bias. Where an unreliable publisher reduces the odds of an objective article by between $\mathbf{0.34}-\mathbf{0.68}$, moving to a left or right political bias does so by between $\mathbf{0.24}-\mathbf{0.25}$. In other words, both reliable and unreliable publishers produce subjective, potentially unverifiable articles, especially when publishing from a biased standpoint. The full model, including estimated marginal medians, and prompt specification, can be found in Appendix \ref{app:additional_results}. We also compare the agreement with \texttt{ChatGPT-4o}, which seems to lean towards an even greater subjectivity propensity.

\paragraph{Manual Annotation} To complement the automated subjectivity analysis, and to verify the alignment of article- and publisher-level labels, we manually annotated a small subset of articles. Specifically, we took 362 articles sampled from the subjectivity annotated subset, stratified over publisher and subjectivity level. Then we check whether the article---in isolation---violates common journalistic norms and practices.

For \unreliable publishers $\mathbf{43.50}\%$ $(36.62\%-50.37\%)$\footnote{The brackets represent the $95\%$ Agresti-Coull binomial proportion confidence intervals \cite{agrestiApproximateBetterExact1998}} of articles were clear cases of non-credible news. The proportion of non-credible articles differs substantially between different publishers, with some \unreliable publishers mixing innocuous articles or clear opinion pieces on general topics with misinformation on specific ones. In \reliable publishers, we deem $\mathbf{8.20}\%$ $(4.07\%-12.32\%)$ of articles to be non-credible. Practically all these cases come from hyper-partisan articles, rather than instances of clear misinformation. Overall, we find that the odds of a non-credible article being published by an unreliable publisher are $\mathbf{8.62}$ times higher than for a reliable publisher.

\begin{figure}[t]
  \centering
  \includegraphics[width=1.0\textwidth]{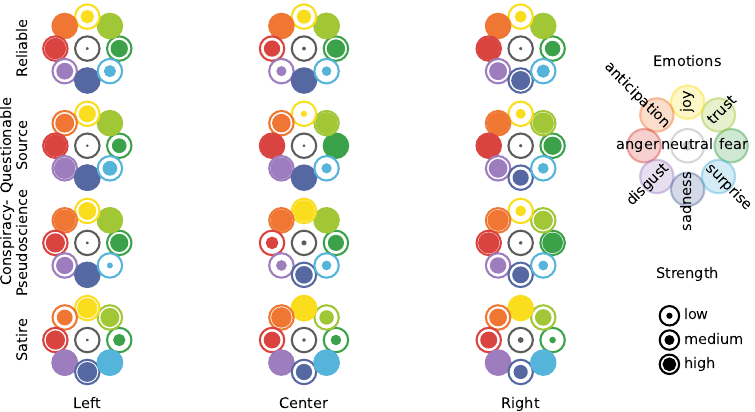}
  \caption{The association of emotions in articles with different publisher categories, as measured using pointwise mutual information. Each circle represents one of Plutchik's 8 emotions, along with neutrality in their center. The area of a circle represents the strength of the association of the emotion with that publisher class (as measured using PPMI), relative to the maximal association found. The legend provides each emotion's color and location in the color wheel.}
  \label{fig:emotion_propensity}
\end{figure}

\paragraph{Emotion analysis} Another property we annotate for is emotion. Affective language in journalistic texts has long been understudied, despite emotion and its effect playing an increasingly important role in the modern media landscape \cite{koivunenEmotiveEvaluativeEpistemic2021}. It is especially prevalent in unreliable news, and is used to both persuade readers and incite sharing \cite{alba-juezEmotionLiesBullshit2019}. While the persuasiveness of emotional language in fake news is a matter of debate \cite{martelRelianceEmotionPromotes2020, phillipsEmotionalLanguageReduces2024}, with prior work showing that high affective state in people after ingesting misinformation being associated with both increased susceptibility \cite{martelRelianceEmotionPromotes2020, bagoEmotionMayPredict2022} and skepticism \cite{hornerEmotionsUnexploredFuel2021, luhringEmotionsMisinformationStudies2024}. From a computational perspective, however, combining emotion detection with misinformation detection has shown some promise \cite{ghanemEmotionalAnalysisFalse2020, zhangMiningDualEmotion2021}. Especially low valence emotions like anger, sadness, anxiety, surprise, and fear are believed to be prevalent in misinformation texts \cite{liuEmotionDetectionMisinformation2024}.

We use a similar setup as above to annotate articles with one of eight Plutchik basic emotions \cite{plutchikChapter1GENERAL1980} and neutrality, a common emotion model used for annotation in NLP \cite{bostanAnalysisAnnotatedCorpora2018}.

Figure \ref{fig:emotion_propensity} shows the association of different emotions with different publisher classes. Visually, these results largely mirror the objectivity analysis; both reliable and unreliable publishers use emotional language, with political bias being a more important determinant of affect than reliability. The most notable difference in association is with `Neutral' and Center-Reliable publishers; relative to other publisher categories, neutral writing occurs relatively often for this publisher class. Low valence emotions like `Anger', `Disgust' and `Sadness' are prevalent throughout, but are especially associated with more politically biased, less reliable publishers. Especially `Satire' publishers seem to be characterized by relatively high amounts of `Joy' and `Surprise'. `Anticipation' is highly prevalent in articles discussing expected future events, which appears to occur regularly, across all publisher classes.

Overall, however, much like subjectivity, there exists a good balance of emotions across the different publisher classes. While there is some correlation between emotion or subjectivity with publisher reliability, these appear to be insufficient to function as a shortcut for misinformation prediction.

While we partially corroborate the finding that different emotions are present (or at least, in different proportions) across publisher classes, ultimately there exists a good balance of emotions across the different publisher classes. Much like subjectivity, simply using the emotions present in an article to determine whether it comes from a reliable or unreliable publisher is likely insufficient.

As argued in Section \ref{sec:introduction}, to reliably estimate the generalization gap of misinformation detectors, it is crucial to have access to a naturalistic corpus of misinformation, which is representative in terms of the diversity contained within. With this analysis, we have shown that for the properties of emotion and subjectivity, this diversity is present, with subjective and emotional language being present in articles from both reliable and unreliable publishers.

\section{Conclusion}
\label{sec:conclusion}

In the present state of the art, automated misinformation detectors cannot be safely or reliably deployed. While impressive performance is often reported, time and again, papers show that in more realistic settings, where out-of-distribution generalization is required, these models fail. In part, this is inherent to the problem of misinformation detection; as \citet{yeeLimitsMachineLearning2025} argues, the informational norms in any community is continually evolving, and any equilibrium is transient. The nonstationary nature of misinformation is, and likely always will be, difficult to model.

One source of this brittleness to distributional shifts, as substantial empirical evidence has shown, is a failure of misinformation datasets to adequately simulate misinformation detection scenarios. Due to the prohibitively expensive cost of procuring misinformation labels, practically all surveyed misinformation datasets have had to make unrealistic assumptions, which introduce undesirable biases.

To accommodate recommended misinformation evaluation practices, and thereby enable the development of \textit{generalizable} misinformation models, this article introduces \texttt{misinfo-general}. It consists of a benchmark built on top of a cleaned, weakly supervised corpus of online articles, which have rich article- and publisher-level metadata, and an operationalization of various generalization axes. We have shown that this dataset is challenging for a common class of misinformation detection models, and especially so when generalization to unseen forms of content or publishers is required.

The metadata annotations enable us to further analyze the determinants of model performance. We find large discrepancies across political biases and misinformation forms, but have also shown that increased diversity has a positive effect on generalization.

While publisher-level labelling introduces noise, we believe the increased scale, diversity, and affluence of metadata make up for this. The first two properties can enable OoD generalization or robustness, whereas the latter enables evaluation, analysis and potentially generalization-aware training.

We make the dataset publicly available, and hope it will serve as a resource for OoD generalization-focused training and evaluation. While the implementation of trustworthy automated misinformation detectors remains out of reach, we hope that this dataset at least makes evaluating and diagnosing generalization capacities of misinformation detection models easy enough for wide-spread academic adoption.


\subsection{Extending \texttt{misinfo-general}}

One of the central claims in this article is that online content and misinformation tends to change, rapidly and comprehensively. While `misinfo-general` represents a large and diverse corpus, useful for pre-training misinformation detectors and evaluating generalization abilities of trained models, it is inevitable that this collection will become misaligned with future forms of misinformation. For example, with the latest articles coming from the end of 2022, it likely misses the newly emerging category of LLM generated (mis)information. Another reason for updating the dataset is to avoid data leakage; new NLP models will likely be trained on collections that overlap with \texttt{misinfo-general}, and can include hyper-textual reference to \texttt{misinfo-general} content (e.g. fact-checking articles). However, we are confident that the collection can be relatively easily maintained and updated.

With the accompanying metadata database, it is trivial to extract information on the publishers already in the dataset, allowing scraping of future content, and these are all linked to the MBFC front-end, which can be used to track labels. Incorporating article-level label providers, which the surveyed datasets in Appendix \ref{app:survey} Table \ref{tab:misinfo_datasets} shows are becoming more accessible, could allow for blending the noisy, but pragmatic distant labelling with high-quality, high-cost article-level labels.

\section{Limitations \& Ethical Considerations}
\label{sec:limitations_ethics}

While the dataset includes a diverse set of publishers, events and topics, ultimately the publisher metadata comes from a single source. The information provided by MBFC assumes a narrow, US-centric world-view. This is especially prominent when discussing foreign publishers from nations with geopolitical ambitions at odds with the U.S. As such, the metadata provides limited nuance, providing only a single perspective of an inherently subjective assessment. It is expected that across cultural backgrounds, the publisher information is bound to change.

In general, the news included in the dataset is US-centric, with the vast majority of publishers being American, producing articles for an American audience. This is exacerbated by us excluding all non-English articles. This means cross-lingual or cross-cultural generalization cannot be evaluated.

That said, we do discuss one weak form of cross-cultural transfer. Besides prioritizing different events, the primary distinction between the political ideologies discussed in Section \ref{sec:Generalization-forms}, are their differing norms and values. As such, the poor political bias generalization bodes ill for the more general cross-cultural generalization tasks.

While much previous work has shown incorporating non-text modalities in the classification pipeline benefits classification \cite{alamSurveyMultimodalDisinformation2022, xiaoChallengesMachineLearning2024}, in its current form, \texttt{misinfo-general} does not include non-text modalities.

In the case of social media context, all references to such content was removed. We consider such content inherently Personally Identifiable Information (PII), and their use in misinformation detection is fraught with ethical problems \cite{mishraModelingUsersOnline2021}.

Embedded images and videos were also not included. This data was not available in the progenitor datasets. This excludes a large and important class of misinformation detectors, which can leverage the interplay of text and non-text context. Incorporating these context would make for valuable future work, allowing for models that more closely emulate the decision process used by human misinformation annotators. At present, however, this lies outside the scope of this project.

In general, the deployment of misinformation detectors comes with legal and ethical issues. Pre-emptive moderating of communication, which is typically the implicit goal of automated misinformation classifiers, is in essence a prior restraint on speech, regardless of the accuracy or OoD robustness of the model \cite{llansoNoAmountAI2020}. While OoD robustness mitigates the propensity of such human rights violations \cite{tobiEpistemicCompassOnline2024}, it cannot remove it entirely.


\subsection{Dataset Access and Licensing}
\label{sec:limitations_ethics_dataset_access}

We aim to make \texttt{misinfo-general} as easy to use as possible, but have had to make some restrictions. The dataset contains texts that are toxic, hateful, or otherwise harmful to society if disseminated. The dataset itself or any derivative formats of it, like language models, should not be released for non-research purposes.

The NELA corpora were initially released under a \href{https://creativecommons.org/publicdomain/zero/1.0/deed.en}{CC0 1.0}\footnote{\url{https://creativecommons.org/publicdomain/zero/1.0/deed.en}}, essentially being released to the public domain. From January 1st, 2024, the NELA authors have de-accessioned their repository. Upon request, the authors note their desire to restrict usage to non-commercial research.

Given the potentially harmful content, and our colleagues wishes, we (re-)release our dataset under a more restrictive \href{https://creativecommons.org/licenses/by-nc-sa/4.0/deed.en}{CC BY-NC-SA 4.0}\footnote{\url{https://creativecommons.org/licenses/by-nc-sa/4.0/deed.en}} license. This allows for redistribution and adaption as necessary for academic research, while preventing commercial use-cases and requiring adaptions to maintain these restrictions. To circumvent copyright of the original texts, we have extended the effort made by the original NELA authors, and have `poisoned' all texts with special tokens.

We have released \texttt{misinfo-general} through two media that allow for restricted access. Specifically, we use Harvard's Dataverse (which implements per-file access restrictions) and HuggingFace's Dataset Hub (which implements repository-level gating). We plan to review access applications manually, limiting use-cases to academic research only.


\clearpage

\appendix
\counterwithin{figure}{section}
\counterwithin{table}{section}
\begin{appendices}

  \section{Misinformation Dataset Survey}
  \label{app:survey}

  \input{tables/datasets_survey}

  \clearpage

  \section{\texttt{misinfo-general} Processing \& Statistics}
  \label{app:dataset_processing}

  \begin{table}[t]
    \centering
    \include{tables/dataset_sizes}
    \caption{Size of the datasets, in terms of millions of articles, after each step of cleaning, per year. The lower percentage gives the step-to-step reduction in size. The 'Total' column computes reduction relative to 'Original'.}
    \label{tab:dataset_sizes}
  \end{table}

  We downloaded the initial NELA corpora from the Harvard Dataverse, under a \href{https://creativecommons.org/publicdomain/zero/1.0/deed.en}{CC0 1.0 license}\footnote{\url{https://creativecommons.org/publicdomain/zero/1.0/deed.en}}. The corpora have since been de-accessioned, and can longer be downloaded. We expand on this in Section \ref{sec:limitations_ethics_dataset_access}.

  \subsection{Re-Labelling}
  \label{app:dataset_relabelling}

  As a first processing step, we relabelled all publishers. This was done to 1) attribute articles to their original publisher (where possible), 2) ensure publisher information was up-to-date (MBFC had expanded their catalogue considerably) and 3) to mitigate the effects of publishers that might interfere with the learning signal.

  An important class of publishers that belong under that third point are \textit{aggregation sites}. Such sites either do not produce original content, or intersperse articles from (usually more reputable) other sources through their content. While the collection of articles as a whole might express some editorial bias, for the most part, these sorts of publisher introduce noise into an already noisy labelling scheme.

  We manually re-mapped all URL domains to a set of publishers consistent across years, excluding all news aggregation platforms and social media sites (see Tables \ref{tab:news_aggregators} and \ref{tab:banned_domains}). It should be noted that the 2018 edition of NELA did not contain URLs, making relabelling in this manner impossible.

  Table \ref{tab:publisher_overlap} shows the amount of publisher overlap exists between different dataset years.

  \clearpage
  \begin{table}[t]
    \centering
    \begin{minipage}{0.40\textwidth}
      \centering
      \caption{Publishers removed for being news aggregation sites, with article counts post de-aggregation.}
      \label{tab:news_aggregators}
      \include{tables/banned_publishers}
    \end{minipage}
    \\[1em]
    \begin{minipage}{0.40\textwidth}
      \centering
      \caption{Publishers removed for being social media sites, with article counts post de-aggregation.}
      \label{tab:banned_domains}
      \include{tables/banned_domains}
    \end{minipage}
    \begin{minipage}{1.0\textwidth}
      \caption{Frequent duplicated articles.}
      \label{tab:frequent_duplicates}
      \include{tables/frequent_duplicates}
    \end{minipage}
  \end{table}
  \clearpage

  \subsection{Deduplication}
  \label{app:dataset_deduplication}

  Models trained on this dataset have to (implicitly) learn a mapping from an article to its publisher's class. As a result, any instances of duplicate content unique to a publisher or publisher class likely induces overfit. The model need only memorize those cases to make a prediction, independent of the article content.

  To minimize the effect of these duplicates, we apply several stages of deduplication.
  \begin{enumerate}
    \item we only keep the first (as determined by publication date) instance of exact duplicates for each publisher. Besides plagiarism, the most common duplicates are due to errors in scraping or parsing, resulting in artifacts (see Table \ref{tab:frequent_duplicates})
    \item we remove all articles with duplicate titles or URLs, across the entire dataset. This primarily removes articles that are updated at a later stage (e.g., live-blogs or summaries), or result from URL re-directions
    \item we remove all articles if more than 5\% of its sentences are duplicates from the same publisher
  \end{enumerate}

  The former two (`Exact Deduplication' in Table \ref{tab:dataset_sizes}) can handle document-level duplicates, like plagiarized articles or updated blog-posts, but miss near duplicates. The latter should filter out near duplicates, like lightly edited documents, (`Near Deduplication' in Table \ref{tab:dataset_sizes}).

  An added benefit of these deduplication steps is the identification of common scraping errors. In fact, we found this to be the most frequent form of duplication. We list some common texts in Table \ref{tab:frequent_duplicates}. Since the majority of such errors were automated responses unique to individual publishers, we could filter these out with relative ease using the described `Exact Deduplication' approaches.

  Ultimately, the separate deduplication steps together resulted in removing 3.1M articles, comprising roughly $\approx 22\%$ the dataset after relabelling.

  \subsection{Cleaning \& Filtering}
  \label{app:cleaning_filtering}

  Many of the included articles either include too many non-semantically relevant tokens or are otherwise malformed. This is aggravated by the NELA authors 'poisoning' the dataset (from 2019 onwards) with repeated `@' tokens to avoid copyright infringement. We do our best to clean the article texts, and remove those with no discernible semantic information.

  We normalize all punctuation and remove any embedded URLs, HTML, or Markdown markup. SpaCy's \cite{montaniExplosionSpaCyV3722023} \texttt{en\_core\_web\_sm} was used to annotate all tokens in the corpora. We identify self-references as named entities with a large longest common substring relative to the publisher's name. We also mask any sentences which SpaCy flags as being part of an email, URL or twitter handle. Finally, we standardize the NELA copyright poisoning, applying it to all dataset years equally.

  \begin{table}[t]
    \caption{Filtering rules.}
    \label{tab:rules}
    \centering
    \include{tables/rules}
  \end{table}

  All-in-all, this introduces 4 new special tokens:  \texttt{<copyright>} replacing NELA's repeated \texttt{@} tokens, \texttt{<twitter>} for X (formerly known as Twitter) handles, \texttt{<url>} for any URLs and \texttt{<selfref>} for any self-references.

  To ensure all articles contained enough grammatical text to reasonably classify, we removed any article which did not abide by the rules delineated in Table \ref{tab:rules}. As a final step, we use Lingua \cite{stahlPemistahlLinguapy2024} to filter out any non-English texts.

  Despite removing almost half of the articles, the dataset retains a level of 'noise' customary to data sourced from the internet. This is inherent to the domain, and further cleaning might negatively affect the realism of the benchmark. One type of noise that could interfere with learning is the presence of stylistically unique substrings identifying publishers. For example, articles from the same publisher tend to contain similar by-lines, attribution messages, or donation requests.

  \subsection{Event \& Topic Clustering}
  \label{app:topic_clustering}

  To generate the metadata necessary for the `Topic' generalization form (see Section \ref{sec:Generalization-forms} or Appendix \ref{app:generalisation_forms}), we opted for a bottom-up approach. This involved first clustering the dataset into thousands of fine-grained events, before clustering the event clusters into overarching topics.

  To achieve this, we used a heavily modified variant of \texttt{BERTopic} \cite{grootendorst2022bertopic} with a \texttt{gte-large}\footnote{\href{https://huggingface.co/thenlper/gte-large}{https://huggingface.co/thenlper/gte-large}} \cite{liGeneralTextEmbeddings2023} backbone.

  After embedding an entire year of the dataset, we first reduce their dimensionality using mini-batched PCA, and whiten the data. We construct a vocabulary over all the documents, by aggregating mini-batches of vocabularies. We then apply \texttt{UMAP} dimensionality reduction, using the PCA solution as initialization, and the cosine distance as the distance metric. Clustering was performed via \texttt{HDBScan} to mini-batches of the embeddings, assigning all articles to their nearest event cluster. Using the dataset-wide vocabulary, we generate a single TF-IDF representation matrix.

  After mini-batching, we generate a hierarchical clustering on top of the event-based TF-IDF matrix. We merge any events with a distance below 0.8 to mitigate the effect of mini-batching, and re-construct the TF-IDF representations. We deem these the final event representations. Table \ref{tab:dataset_counts} provides the total number of events and the average number of articles contained within each.

  Finally, we apply a round of spectral clustering to these representations, compressing these events into 10 groups, which we deem topics. The numbers of events in each topic varies (see Table \ref{tab:dataset_counts}).

  \begin{figure}[t]
    \includegraphics[width=\textwidth]{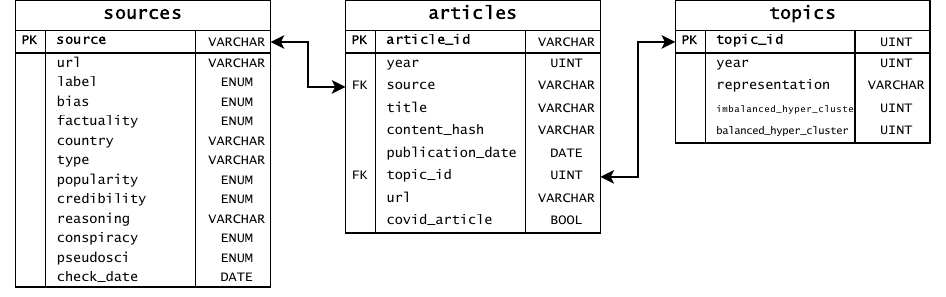}
    \caption{Schema for the generated metadata database. 'PK' indicates primary key, 'FK' denotes a foreign key relationship.}
    \label{fig:metadata_schema}
  \end{figure}

  \subsection{Metadata}
  \label{app:metadata}

  Article content is stored in \texttt{arrow} files, accessed using HuggingFace's \texttt{datasets} library \cite{lhoest-etal-2021-datasets}. We store all generated metadata in a \texttt{duckdb} database \cite{duckdb}. Figure \ref{fig:metadata_schema} depicts the metadata schema. Table 'Sources' provides information sourced from MBFC on different publishers, whereas tables 'Articles' and 'Topics' provides information on the produced articles and events/topics, respectively. The articles are all given unique article identifiers. This ensures we can quickly generate relevant splits of the dataset and link these to article-level predictions.

  \subsection{Publisher-level Metadata}
  \label{app:publisher_metadata}

  All publisher metadata is stored in the 'Sources' table. Each source is identified by a unique name, linked in a 1-to-many relationship to articles in the 'Articles' table. The \texttt{url} column provides a link to the MBFC page with the source metadata. The \texttt{label} column provides the MBFC label, also shown in the lower rows of Table \ref{tab:dataset_counts}. In column \texttt{bias}, the political bias (one of 'Extreme Left', 'Left', 'Left-Center', 'Least Biased', 'Pro-Science', 'Right-Center', 'Right' or 'Extreme Right') of the publisher is provided. The \texttt{factuality} column provides an ordinal value for the propensity of a publisher to report factual news. Columns \texttt{country} and \texttt{type} provide information about the country of origin of the publisher, and their primary form of publication (e.g., TV, blog posts), respectively. In column \texttt{popularity}, the average number of visits to the publisher's main website is provided as an ordinal categorical value. \texttt{credibility} provides an overall assessment of the publishers' credibility, aggregating all other variables into a single score. Finally, the \texttt{conspiracy} and \texttt{pseudosci} columns provide the `strength' of the conspiracy or pseudo-science source. The larger this value is, the more these sources deviate from the public or scientific consensus. This is only provided for publishers labelled as `Conspiracy-Pseudoscience'.

  \begin{table}[t]
    \centering
    \caption{Publishers whose MBFC label changed substantively, from initial data collection (2017) to the publication of \texttt{misinfo-general} (Oct. 2024).}
    \label{tab:publisher_label_changes}
    \include{tables/publisher_changed_labels}
  \end{table}

  The MBFC label and the metadata of publishers is subject to change, as additional information is made available or corrections are processed. We list the 20 publishers whose labels changed substantively during the data collection period in Table \ref{tab:publisher_label_changes}. In most instances, the label changed from \reliable to \unreliable due to additional fact check failures being incorporated into the MBFC database (12/20). The inverse also occurs, where publishers improve their editorial practices or correct articles that failed fact checks (5/20). Most metadata changes correspond to relatively minor changes in political bias (e.g., Left-Center to Right-Center), and do not alter the analyses presented in this article.

  \subsection{Article Statistics}
  \label{app:article_statistics}

  \begin{table}[t]
    \centering
    \caption{Various statistics on the dataset and labels. \textsc{Article Counts} provides the total number of articles in each dataset year. \textsc{Label Proportion} provides the relative occurrence of reliable vs. unreliable articles, whereas \textsc{Publishers} provides the number of such publishers present. Section \textsc{Truncated Token Counts} provides the mean and the 25, 50, 75th quantiles of the number of tokens per article. Note that is after truncation at 512; the raw articles tend to be much longer. \textsc{Event Clustering} provides the number and average size (in articles) of events. We also provide the number of events belonging to the smallest and largest topics. Sections \textsc{Reliable Labels} and \textsc{Unreliable Labels} provides the proportion of each MBFC label in the dataset, split into publisher categories.}
    \label{tab:dataset_counts}
    \include{tables/dataset_counts}
  \end{table}

  We present various statistics at the article level for each iteration of the final \texttt{misinfo-general} dataset in Table \ref{tab:dataset_counts}. This includes the number of articles, the average number of tokens per article (post truncation), the number of events and the size of topics, and finally the label proportions, both in aggregated form as \reliable or \unreliable, but also per publisher category. In total, the dataset comprises some 2B tokens once truncated. Without truncation, this is likely substantially higher.

  In Table \ref{tab:topics_example} we present descriptions of the various topics in the 2020 split of \texttt{misinfo-general}. These topics are latent, and automatically generated from the inter-event distance matrix. The space in which the articles were clustered into events is high-dimensional, and each cluster was assumed to be non-convex. As a result, it is difficult to find an 'average' representation of any event, and even more difficult to find one for a topic (i.e., a cluster of clusters). Instead, we derived topic descriptions by sampling events from each topic, and using the \texttt{BERTopic} generated event representation.

  \begin{table*}[t]
    \centering
    \caption{An example of the topics present in the 2020 split of \texttt{misinfo-general}. As described in Section \ref{sec:Generalization-forms}, to form the OoD set, the $k$ smallest topics are chosen, such collectively they (approximately) comprise 20\% of all articles.}
    \label{tab:topics_example}
    \input{tables/topic_example}
  \end{table*}

  \subsection{Publisher Statistics}
  \label{app:publisher_statistics}

\begin{figure*}[t]
    \centering
    \includegraphics[width=\textwidth]{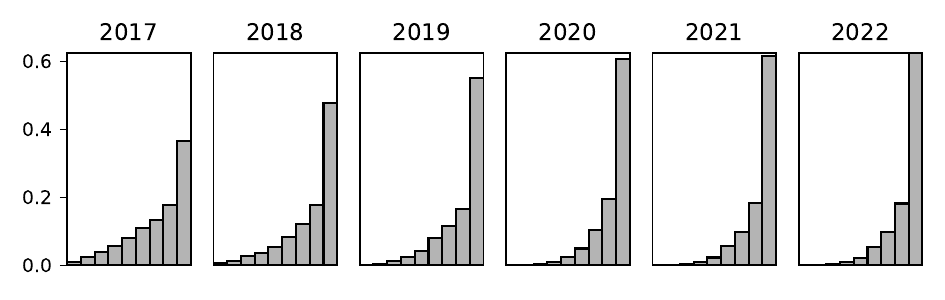}
    \caption{The authorship of articles in the dataset, aggregated into deciles. The right most column of each panel represents the top 10\% most prolific publishers, the left most the bottom 90\%.}
    \label{fig:authorship_deciles}
  \end{figure*}

  Across all dataset iterations, the vast majority of articles are written by a small minority of prolific publishers. This is shown in Figure \ref{fig:authorship_deciles}. The most prolific publishers have authored between $40$-$60$\% of articles within a year. The next deciles manage only $20$\%, then $10$-$15$\%, etc., with the smallest publishers authoring only a few articles. This effect seems to get more drastic for the later dataset years. \\

  As discussed in Section \ref{sec:results} this can obscure poor performance when using article-level evaluation metrics. In such cases, the average performance score will tend to be dominated by the performance of the largest publishers, instead of the body of publishers as a whole. As shown in Figure \ref{fig:determinants} and Tables \ref{tab:determinants_coefficients} and \ref{tab:determinants_coefficients_topic}, performance on these less mainstream publishers tends to be substantially worse. As a result, the OoD performance metrics in Table \ref{tab:generalisation_forms} likely underestimate the generalization gap.

  This dataset property means publisher-level analyses are necessary, in addition to the far more common article-level performance measures. That said, prominent misinformation datasets tend to sample from a far more a smaller and homogenous set of publishers, with typically more balanced authorship proportions.

  Another noteworthy statistic is the amount of overlap between publishers across the dataset years. While the dataset as a whole contains $488$ distinct publishers, as can be seen in Table \ref{tab:dataset_counts}, each dataset year contains fewer total publishers. Furthermore, the distribution of those publishers (in terms of article counts) shifts across the years. This is partially due to a changing collection methodology used by the NELA authors.

  \begin{wraptable}{r}{0.50\textwidth}
    \vspace{-1.3em}
    \centering
    \include{tables/publisher_overlap}
    \caption{Overlap of publishers between years.}
    \label{tab:publisher_overlap}
  \end{wraptable}

  As a result, the different dataset years have different degrees of overlap between each other. Table \ref{tab:publisher_overlap} shows the publisher intersection-over-union overlaps between the $6$ different dataset years. The largest block of overlap is between the years 2020-2022, with overlaps across datasets being well above $0.80$. The IoU scores to other dataset years are far lower, with 2017 having particularly low overlaps.

  \clearpage

  \section{Additional Information on Generalization Axes and Splits}
  \label{app:generalisation_forms}


  \subsubsection*{Uniform} For this generalization axis, we employ the industry standard method of stratified and shuffled train/validation/test splits. We maintain the ratios 80/10/20, respectively, throughout. This is meant to form the baseline set of values, both in terms of ID--OoD articles performance, but also the expected delta between those.

  \subsubsection*{Time} Misinformation changes quickly, but models have been found to be brittle to articles outside the time-span of the training data \cite{bozarthBetterPerformanceEvaluation2020, horneRobustFakeNews2020}. We train models on a dataset from a single year, and evaluate performance on all other years. To avoid conflating `Time' and `Publisher' results, we only evaluate on articles from publishers seen in the training set. The proportion of publishers--and as a result, the reliable/unreliable proportions--does change drastically across dataset years.

  \subsubsection*{Event} We define events as occurrences that span news articles with a definite time-span and sudden occurrence. Novel events spawn articles with a markedly different vocabulary, introducing words, names, or terms not yet available to the model. Prior work has shown models suffer when evaluated on unseen events \cite{leeUnifyingMisinformationDetection2021, dingMetaDetectorMetaEvent2022}. The dataset contains a multitude of such events, but we focus specifically on the COVID-19 pandemic; an event that particularly notable for its rapid onset and the volume of misinformation it spawned \cite{braddInfodemic2024}. Models are trained on a subset of articles \textit{not} containing COVID-19 keywords, and evaluated on ones that do. The article discovery-process is outlined in more detail in Appendix \ref{app:event_discovery}.

  \subsubsection*{Topic} We define topics to be relatively static collections of events, wherein the style or publishers' opinions change little across years or decades. We discover such topics automatically, bottom-up. Specifically, we apply a heavily modified variant of the \texttt{BERTopic}\cite{grootendorst2022bertopic} algorithm to discover the many thousands of events that occur in each year of the dataset. Each event is textually represented by a TF-IDF vector. To group events into topics, we apply a second round of spectral clustering on the adjacency matrix induced by the inter-event cosine distance matrix. The number of articles contained by a topic differs substantially (see Table \ref{tab:dataset_counts}). We reserve the $n$ smallest topics, which collectively contain roughly 20\% of all articles for testing, and train on articles from all other topics.

  \subsubsection*{Publisher} All scraped articles are annotated with the news website, outlet, or publisher that produced the article (see Figure \ref{fig:metadata_schema}). In general, news publishers have some editorial bias or stance, making their corpus of articles distinctly different from that of other publishers. Prior work has found misinformation classifiers to be particularly sensitive to changes in publisher \cite{barron-cedenoProppyOrganizingNews2019,bozarthBetterPerformanceEvaluation2020}. We specifically test for this by reserving the $n$ smallest publishers, which collectively contain roughly 20\% of all articles for the test set, and train on articles from all other publishers. We stratify this splitting across political bias, to ensure each political bias is represented equally in both the train and test sets.

  \subsubsection*{Political Bias} Separate from editorial bias, publishers tend to exhibit a political bias as well. MBFC has annotated all publishers on a left-right political bias continuum. We map all `Extreme Left' and `Left' publishers to a coarser `Left' group, and vice versa for the `Right'-side of the political spectrum. All other political biases (`Center-Left', `Center-Right', `Least Biased' and `Pro-Science') were mapped to `Center' bias. We reserve either the `Left' or `Right' group for evaluation, and train on the collection of articles from the opposite political bias, along with all `Center' biased publishers.

  \subsubsection*{Misinformation Type} Misinformation presents in various forms. MBFC divides unreliable publishers in three categories: `Questionable-Source' (publishers that exhibit extreme bias or propaganda), `Conspiracy-Pseudoscience' (publisher aligning with known conspiracies or pseudo-scientific practices) and `Satire' (publishers whose content is purposefully false for comedic effect). To test this form of generalization, we reserve either `Questionable-Source' or `Conspiracy-Pseudoscience' labels for evaluation, and train on all other articles. The test set includes a stratified sample of reliable sources as well.


  \subsection{Identifying COVID-19 Articles}
  \label{app:event_discovery}
  \begin{figure*}[!t]
    \centering
    \includegraphics[width=\textwidth]{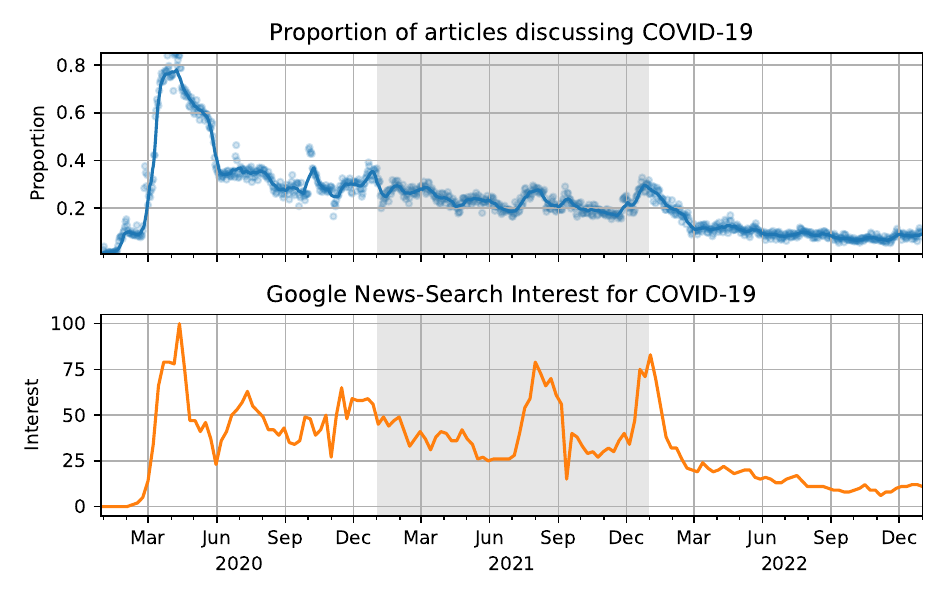}
    \caption{The top figure shows the proportion of articles published that day which discuss the COVID-19 pandemic. The solid line provides a LOESS smoothed trend line. The bottom figure provides a Google search interest for topics that fall under \href{https://trends.google.com/trends/explore?date=today\%205-y\&geo=US\&q=\%2Fg\%2F11j2cc_qll}{Coronavirus disease 2019}. The different years (2020, 2021, 2022) are displayed as banded vertical columns.}
    \label{fig:covid_vs_google_news}
  \end{figure*}

  For `Event' splitting, we focus on a single event present over the latter 3 dataset years: the COVID-19 pandemic. We first derive a set of COVID-19 keywords (e.g., `sars-cov-2', `lockdown', `mask'), and combine it with the set of keywords defined by \cite{gruppiNELAGT2020LargeMultiLabelled2021}. We include all articles from 2020-2022 that include any of these terms in the held-out test set. While this is bound to induce a large number of false positives, it ensures no COVID-19 related terms contaminate the models' learned vocabulary. In Figure \ref{fig:covid_vs_google_news} we display the correlation between the found number of articles and the amount of Google search volume.

  \subsection{Test Set Correlations}
  \label{app:test_test_correlations}

  \begin{table}[!t]
    \centering
    \include{tables/publisher_topic_overlap}
    \caption{The amount of \texttt{article\_id} overlap between the `Publisher' and `Topic' test sets. Column `IoU' gives the intersection over the union between the two held-out test sets. Column `Over Random' compares this to the achieved value for $10000$ simulated draws from a uniform distribution of the same sizes as the actual test set sizes (given in the last two columns).}
    \label{tab:publisher_topic_overlap}
  \end{table}

  Publishers have a tendency to prioritize different topics. This is especially prevalent for less mainstream publishers, which have far fewer resources to spend on the breadth of their reporting. It is therefore plausible that the `Publisher' and `Topic' splits have some overlap.

  We test for this by comparing the amount of article level overlap in the held-out test sets of the `Publisher' and `Topic' to random draws of the same size from uncorrelated uniform distributions. The larger this overlap, the more correlated the two test sets are. We display the results of our test in Table \ref{tab:publisher_topic_overlap}. While we find more overlap than random draws, the effect is not large. Even in the most egregious case, there is only a $0.04$ difference in absolute IoU. As such, we believe the two different test sets measure distinct generalization aspects.

  \clearpage

  \section{Training Details}
  \label{app:training_details}

  \begin{wraptable}{r}{0.45\textwidth}
    \vspace{-1.3em}
    \centering
    \include{tables/hyperparameters}
    \caption{An overview of important hyperparameters and their values. These were tuned on the validation set of the `Uniform' split dataset.}
    \label{tab:hyperparameters}
  \end{wraptable}

  While \texttt{DeBERTa-v3} models can theoretically handle infinite sequences, to constrain memory requirements we truncate tokenization after the first $512$ tokens. This allowed for a batch size of $64$ during training, and $512$ during validation. We used \texttt{Adam} with weight decay \cite{kingmaAdamMethodStochastic2015,loshchilovDecoupledWeightDecay2019} as the optimizer.

  Some other important training hyperparameters are listed in Table \ref{tab:hyperparameters}. We set a high learning rate for the classifier and pooler, but a very low value for the embeddings layer to avoid catastrophic forgetting. We employed a polynomial decaying learning rate scheduler, which first linearly warms up, before decaying according to
  \begin{equation*}
    \eta_{t}=\left(\frac{t}{\text{\texttt{max\_steps}}}\right)^{\texttt{power}}\eta_{t-1}
  \end{equation*}
  For each experiment, regardless of the size of the dataset, we allow for a maximum of $3.0e+6$ update steps, evaluating on the validation set every $5$\%. We employ early stopping with patience of 2, through which training usually concludes well before reaching the maximum budget.

  \subsection{Additional Training Details}

  All experiments were conducted on SNELLIUS, a Linux SLURM-based supercomputer. Nodes consist of an Intel Xeon Platinum 8360Y CPU with 18 cores, 2.4 GHz speed, and a single NVIDIA A100 GPU accelerator (40 GiB of HMB2 memory), and 128 GiB of DDR4 memory.

  We use Python 3.11 and PyTorch 2.2.2 built with CUDA 11.8. For training, we make extensive use of utilities implemented in HuggingFace's \texttt{transformers} library: \texttt{transformers} 4.37.2, \texttt{datasets} 2.19.0 and \texttt{accelerate} 0.30.1. All experiments were conducted under random seed 942. For local development, we use Ubuntu 20.04.6 LTS (GNU/Linux 5.15.90.1-microsoft-standard-WSL2 x86\_64).


  \subsection{LLM Inference}
  \label{app:llm-inference}

  To conduct LLM inference, we use Meta's \texttt{llama-3-8B-Instruct} using the following prompt:

  \begin{quote}
    \noindent
    <<SYS>>

    \noindent
    You are a content moderator working with journalistic articles. Your task is to identify articles from unreliable publishers.

    \noindent
    <</SYS>> \\

    \noindent
    Does the following text come from a reliable news publisher?

    \noindent
    Respond with 'yes' or 'no'.

    \noindent
    Article: \texttt{\$\{ARTICLE\}}

    \noindent
    Does this article come from a reliable news publisher? 'yes' or 'no': [/INST]
  \end{quote}

  Here, \texttt{\$\{ARTICLE\}} is replaced by the first $512$ tokens of each article.

  We use the \texttt{pmi\_dc} decision rule, introduced by \citet{holtzmanSurfaceFormCompetition2021}. We use the same prompt without the article tokens as the domain conditional text. Empirically, this performs slightly better across the entire corpus.

  Due to the (very) long articles prevalent in this dataset, using few-shot exemplars was deemed intractable.

  \subsection{Annotating Article Properties}
  \label{app:article_properties}

  \paragraph{Subjectivity}

  \begin{wraptable}{r}{0.45\textwidth}
    \vspace{-1.3em}
    \centering
    \include{tables/subjectivity_annotation_confusion_matrix}
    \caption{A confusion matrix between the subjectivity assessments of \texttt{ChatGPT4o} (rows) and \texttt{ChatGPT4o-mini} (columns). }
    \label{tab:subjectivity_annotation_correspondence}
  \end{wraptable}

  To annotate articles for subjectivity, we used \texttt{ChatGPT4o-mini}, and \texttt{ChatGPT4o}. The former was used to annotate the bulk of articles, and the latter to verify the overall quality of the annotations. We used the following prompt:

  \begin{quote}
    \noindent
    <<SYS>>

    \noindent
    You are a helpful assistant, helping analyse the properties of news articles. Before a final answer, make sure to explain your thinking.

    \noindent
    <</SYS>> \\

    \noindent
    Please classify how objective the following article is.

    \noindent
    Objective articles take a neutral stance on topics, and focus on reporting factual news.

    \noindent
    Subjective articles instead focus on opinions, which are more difficult to verify and can take specific stances for or against topics.

    \noindent
    The title and body are provided. After you provide your reasoning, respond with one of {entirely objective, mostly objective, mixed, mostly subjective, entirely subjective}, and nothing else.

    \noindent
    Article: \texttt{\$\{ARTICLE\}}
  \end{quote}

  Table \ref{tab:subjectivity_annotation_correspondence} gives the classification correspondence between the two models. Overall, we find that the mini model aligns well with the larger model, and that the larger model tends towards \textit{more} subjective annotations, rather than less. As such, it is plausible that the subjectivity within the dataset remains underreported. We include some examples of articles and their subjectivity annotations in \url{https://github.com/ioverho/misinfo-general/tree/main/assets/subjectivity_annotations_examples}.

  \paragraph{Emotion}

  We perform a similar analysis, but now for emotions present in the articles. We use the following prompt, and only use \texttt{ChatGPT4o-mini}:

  \begin{quote}
    \noindent
    <<SYS>>

    \noindent
    You are a helpful assistant, analyzing the properties of news articles. Before a final answer, make sure to explain your thinking.

    \noindent
    <</SYS>> \\

    \noindent
    Please identify the dominant emotions in the following article.

    \noindent
    We will be using Plutchik's 8 basic emotions, along with a label for neutral:
    \begin{enumerate}
      \item Joy: a feeling of happiness, pleasure, or contentment.
      \item Sadness: a feeling of loss, disappointment, or grief
      \item Trust: a sense of safety, security, and connection with others
      \item Disgust: a strong aversion to something unpleasant, often related to taste, smell, or moral judgments
      \item Fear: a response to perceived danger, leading to caution or escape behaviors
      \item Anger: a reaction to perceived threats or injustice
      \item Surprise: a reaction to something unexpected
      \item Anticipation: looking forward to or expecting something, which can bring excitement or anxiety
      \item Neutral: emotional balance, no strong positive or negative emotions are present
    \end{enumerate}

    \noindent
    To respond, please list the emotions present in the article. When labelling the article for the 'Neutral' emotion, please make sure no other emotion is present.

    \noindent
    The article's title and body are provided. After you provide your reasoning, respond with a list of {{joy, sadness, trust, disgust, fear, anger, surprise, anticipation, neutral}}, and nothing else.

    \noindent
    Article: \texttt{\$\{ARTICLE\}}
  \end{quote}

  \begin{figure}
    \centering
    \includegraphics[width=\textwidth]{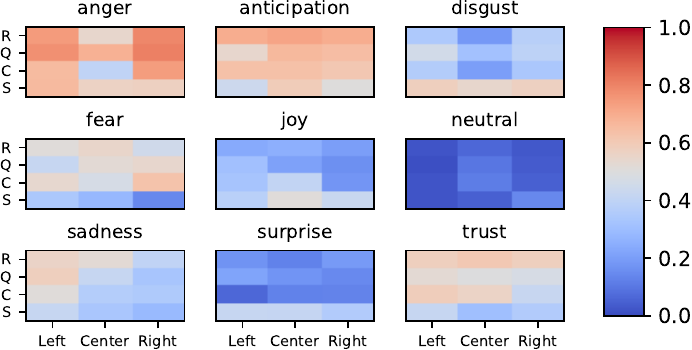}
    \caption{Emotion propensity, like Figure \ref{fig:emotion_propensity}, but now in absolute values, and presented for each emotion in isolation. More intense, red colors signify a higher presence of an emotion, whereas cooler, blue colors the opposite.}
    \label{fig:emotion_propensity_2}
  \end{figure}

  Figure \ref{fig:emotion_propensity_2} presents the same data as in Figure \ref{fig:emotion_propensity}, but now transposed (all publisher class combinations per emotion), and in absolute values. We find emotion overall to be present in all forms of articles and publishers.

  \clearpage

  \section{Additional Results}
  \label{app:additional_results}

  \subsection{Political Bias}
  \label{app:political_bias}

  \begin{figure*}[t]
    \centering
    \includegraphics[width=0.9\textwidth]{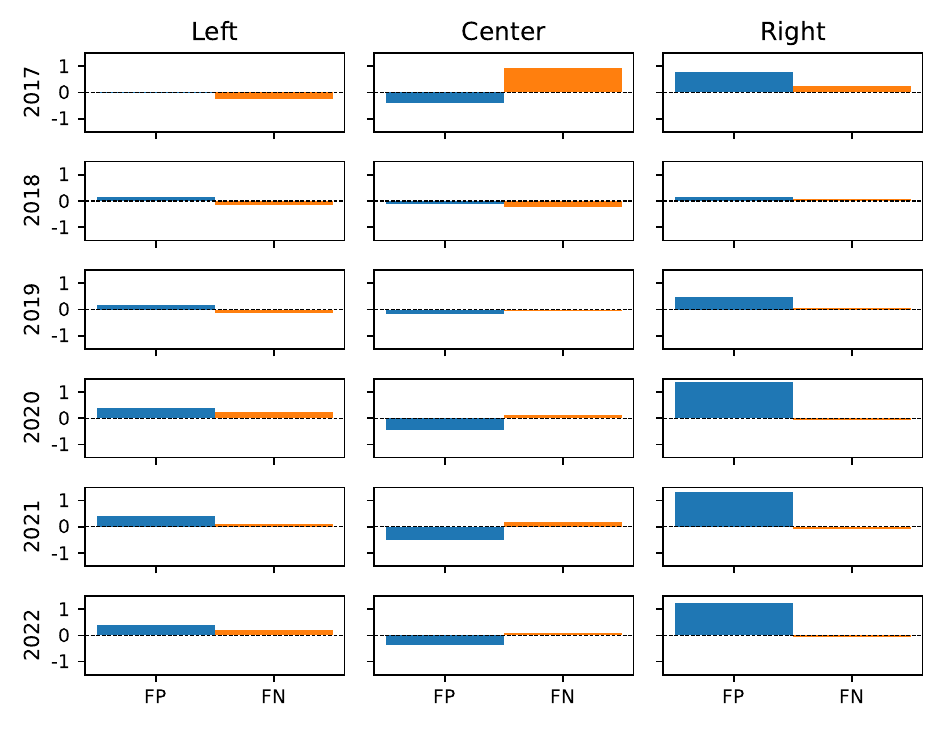}
    \caption{Probabilities of making an error for publishers of a particular label and political bias, relative to the probability of making a mistake for publishers of the same label. FP columns denote the probability of error for reliable publishers, and FN vice versa.}
    \label{fig:model_political_bias}
  \end{figure*}

  To further assess the model's interaction with political bias, we compute the complement of the expected publisher-level accuracy score for a political bias and publisher label (reliable or unreliable), and divide it by the marginal expected score:
  \begin{equation*}
    \frac{1-\mathbb{E}[\text{\textsc{Acc}}_p|\text{bias}(p), y_p]}{1-\mathbb{E}[\text{\textsc{Acc}}_p|y_p]}-1
  \end{equation*}
  essentially giving the shifted exponentiated pointwise mutual information (PMI) of making an error for a particular combination of publisher political bias and label,
  \begin{equation*}
    \frac{p(\hat{y}_{p}\not=y|\text{bias}(p), y)}{p(\hat{y}_{p}\not=y|y)}-1=\exp\left(\text{pmi}\left(\hat{y}_{p}\not=y;\text{bias}(p)|y\right)\right)-1
  \end{equation*}
  The higher this value is, the greater the association between making an error for a particular political bias deviates from the expected value. If an error is made for a reliable publisher, this corresponds to a False Positive (FP) whereas an error for an unreliable publisher would correspond to a False Negative (FN). We introduce the shifting to ensure no bias corresponds with 0, whereas positive and negative bias correspond to positive and negative values, respectively.

  In Figure \ref{fig:model_political_bias} we display various such biases for each iteration of the corpus. As seen in Table \ref{tab:determinants_coefficients}, performance degrades for the more extreme political biases. There is, however, a notable difference between left-biased and right-biased publishers, and especially for the reliable right-biased publishers. The probability of the model generating a False Positive is substantially higher (sometimes near twice as likely) than for reliable leftist or center biased publishers. We conclude that the model tends to confuse articles from reliable right biased publishers with unreliable ones, far beyond what would be considered 'due to chance'.

  On the other hand, the model tends to be overly optimistic for center-biased publishers, with lower FP probabilities and a tendency for slightly higher FN probabilities. An unreliable publisher could therefore escape moderation by limiting its explicit political views. While this is a relatively unlikely example, as most unreliable publishers are unreliable \textit{because} of overt extreme political views, it does showcase an important blind-spot for these models.

  Given the high-stakes nature of misinformation moderation decisions, this indicates the presence of undesirable behavior. In their current state, our misinformation detection models discriminate against publishers of a particular political group.

  \subsection{Publisher Heterogeneity}
  \label{app:publisher_heterogeneity}

  \begin{wraptable}{r}{0.45\textwidth}
    \vspace{-1.3em}
    \centering
    \include{tables/dataset_effect_counts}
    \caption{Mean number of articles and publishers retained in the training set after limiting to the top-n most prolific publishers. Note that the number of articles and publishers deviates from the distributions depicted in Figure \ref{fig:authorship_deciles}. In this case, we additionally condition on the publisher's MBFC label.}
    \label{tab:dataset_effect_counts}
  \end{wraptable}

  In Section \ref{sec:analysis_dataset_effect} we tested for the effect of publisher heterogeneity on generalising to unseen publishers. We operationalized this by re-running the `Publisher' split experiment with only the top-n most prolific publishers for each MBFC label, while leaving the test set untouched.

  This dramatically reduces the amount of variation in publishers, while minimally affecting the total amount of data present. This can be seen in Table \ref{tab:dataset_effect_counts}. Note that these are expressed as percentages, but in absolute terms constitute hundreds of thousands of articles.

  \subsection{Determinants of Publisher-Level Accuracy}
  \label{app:determinants}

  To estimate which factors impact accuracy at the publisher level. We define publisher level accuracy for a publisher $p$ as,
  \begin{equation*}
    \text{\textsc{Acc}}_{p}=\frac{1}{|\mathcal{D}^{(p)}|}\sum_{d\in\mathcal{D}^{(p)}}\mathds{1}(\hat{y}^{(d)}=y^{(p)})
  \end{equation*}
  or put otherwise, the expected rate of correct classification for all documents belonging to that publisher.

  To determine which factors contribute to these scores, we use a multinomial logistic regression model. Specifically, we use \texttt{statsmodels} \cite{seabold2010statsmodels} to estimate the dependent variables as,
  \begin{equation*}
    \sum_{d\in\mathcal{D}^{(p)}} \mathds{1}(\hat{y}^{(d)}=y^{(p)}) + \mathds{1}(\hat{y}^{(d)}\not=y^{(p)}) \\
    =\sigma(\beta_{0}+\sum_{i=1}^K\beta_{i}x_{i})
  \end{equation*}
  We use $\mu^2\cdot (1-\mu)^2$ as the variance function, Pearson's $\chi^2$ as the scale value and the logit as the link function $\sigma$.

  \subsubsection*{Uniform Generalization}

  The fully specified model is displayed in table \ref{tab:determinants_coefficients}, with some important variables' coefficients depicted graphically in Figure \ref{fig:determinants}. The left most column provides natural groupings of the different variables. The interpretation of the coefficient magnitudes should be adjusted to consider the range of possible values for the corresponding variable.

  In the \textit{Generic} category, we include the logarithm of the number of articles present in the training data, and whether the publisher is foreign or not. Both variables are assigned large coefficients. Every 10x increase of labelled articles increases the odds of correct classification by a factor of $1.91$. Foreign publishers tend to be $2.91$ times as easy to classify relative to U.S. based publishers.

  The \textit{Year} category of variables includes a dummy variable for each iteration of the dataset. Here we find relatively little difference, despite statistical significance, with coefficients falling between $0.85$ and $1.15$.

  We further include dummy variables for the different political biases and MBFC labels (assumes center-Reliable to be the default group). We furthermore include an interaction effect between all political biases and MBFC labels. We find that unreliable publishers are substantially more difficult to classify, with all unreliable labels having large negative coefficients. This is especially true for the 'Questionable Source' category of articles, with log-odds shifts of $-1.25$ and $-1.84$ for left and right biased publishers, respectively. Furthermore, corroborating the analysis in Appendix \ref{app:political_bias}, the models perform much more poorly for publishers on the right side of the political spectrum.

  In the \textit{Strength} block of variables, we include MBFC's strength of conspiracy or pseudo-science. All `Conspiracy-Psuedoscience' sources are rated on a scale, with larger values indicating a further deviation from the societal norm. Reliable sources a given a default score of $0$. We find that there exists a slight, but consistent, positive correlation with the \textit{strength} of the pseudo-scientific or conspiratorial claims. In other words, the more those publishers deviate from the status quo, the more easily these are identified.

  Similarly, for the \textit{Factuality Score} block, we include the MBFC score for a publisher's propensity for factual reporting as interaction effects with each MBFC label (excluding `Satire'). Again, we find a slight positive correlation for reliable sources, meaning that the more often such publishers report factual information, the better classification performance. For the unreliable publishers, however, the opposite holds: the less likely to present factual news, the easier the classification. In either case, an 'average' level of factuality tends to correspond to more ambiguous cases, lying closing to the model's decision boundary.

  \subsubsection*{Topic Distance}

  We repeat this analysis, but this time aggregating entities at the event/label level, using the `Topic' split results. The model specification is left as above, with the introduction of a single additional variable: the minimum cosine distance to any event in the training set. The coefficients are provided in Table \ref{tab:determinants_coefficients_topic}.

  While comparison across Tables \ref{tab:determinants_coefficients} and \ref{tab:determinants_coefficients_topic} is not directly possible, as each estimates using different entities and different models, it is encouraging to see similar magnitudes for most variables' coefficients.

  Surprisingly, we observe a very small but positive effect: $\beta_{\text{min\_dist}}=0.15$. This implies that a topic that is as far away as possible from all topics in the training set sees a $1.16$ increased odds of correct classification. While this is a weak effect, it does suggest that events more typical to a certain topic are easier to classify in than those near the topic cluster borders.

  \begin{table*}[p]
    \centering
    \include{tables/determinants_table}
    \caption{Coefficients of the publisher-level determinants model, using the uniformly split models. Range indicates the possible values each variable can take, with $..$ indicating all natural numbers between the extremes (inclusive). $\beta$ provides the logistic parameter, and $\exp(\beta)$ its exponent (i.e. the log-odds ratio). Column 'Std. Err.' provides the standard error, 'p' the corresponding $p$-value and '95\% CI' the confidence interval. In the parameter names, the ':' indicates an interaction term.}
    \label{tab:determinants_coefficients}
  \end{table*}

  \begin{table*}[p]
    \centering
    \include{tables/determinants_table_topic}
    \caption{Idem, but now for the publisher-topic aggregated binomial logistic model. Since the aggregation level differs from that of the model presented in Table \ref{tab:determinants_coefficients}, the coefficients are not directly comparable.}
    \label{tab:determinants_coefficients_topic}
  \end{table*}

\end{appendices}


\newpage

\starttwocolumn

\begin{acknowledgments}
This work was supported by Meta through a Facebook Research Individual Research Project grant, through the project ``Modelling emotion-inducing communication strategies to detect misinformation".

We would like to thank the anonymous reviewers for their time, effort, and insights. Their work has helped improve this article substantially. 

We additionally would like to thank Mauricio Gruppi and Benjamin Horne, along with their colleagues, for making the NELA datasets available and their rapid response to questions.
\end{acknowledgments}

\bibliographystyle{compling}
\bibliography{bib_zotero,custom}

\end{document}

%% file: tables/generalisation_taxonomy.tex
{
\begin{tblr}{
    colspec = {@{}X[1,c]X[1,c]X[3,c]@{}X[3,c]@{}},
    hline{1,2,8} = {0.5pt,solid},
    hline{3,5} = {0.5pt,solid,gray},
    hline{4,6,7} = {0.5pt,solid,gray},
    cell{1}{1} = {r=1, c=2}{c},
    cell{3}{1} = {r=2, c=1}{c},
    cell{5}{1} = {r=3, c=1}{c},
    row{5} = {2.0em},
    }
    \textbf{Generalisation Axis} & & \textbf{In Distribution} & \textbf{Out-of-Distribution} \\
    \textbf{Time}
    & Time
        & \begin{tabular}{ccc}
            CNN & AP & Vox \\
            2018 & 2018 & 2018
            \end{tabular}
        & \begin{tabular}{ccc}
            CNN & AP & Vox \\
            2017 & 2020 & 2019
            \end{tabular}
        \\
    \textbf{Content}
    & Event
        & not COVID-19 events
        & COVID-19 events \\
    & Topic
        & Crime, Sports
        & Elections
        \\
    \textbf{Publisher}
    & \makecell{Publisher\\}
        & \begin{tabular}{@{}ccc@{}}CNN & MSNBC & OANN\end{tabular}
        & \begin{tabular}{@{}ccc@{}}Reuters & AP & True Activist\end{tabular}
        \\
    & \makecell{Political\\ Bias}
        & \begin{tabular}{@{}ccc@{}}
            \scriptsize{AP} & \scriptsize{Reuters} & \scriptsize{Fox News} \\
            \scriptsize{Centre} & \scriptsize{Centre} & \scriptsize{Right}
        \end{tabular}
        & \begin{tabular}{@{}ccc@{}}
            \scriptsize{Vox} & \scriptsize{Daily Beast} & \scriptsize{True Activist} \\
            \scriptsize{Left} & \scriptsize{Left} & \scriptsize{Left}
        \end{tabular}
        \\
    & \makecell{Misinfo\\ Type}
        & \begin{tabular}{@{}ccc@{}}
            \scriptsize{Vox} & \scriptsize{NYT} & \scriptsize{OANN} \\
            \scriptsize{Reliable} & \scriptsize{Reliable} & \scriptsize{Questionable}
        \end{tabular}
        & \begin{tabular}{@{}ccc@{}}
            \scriptsize{MSNBC} & \scriptsize{911Truth} & \scriptsize{Age of Autism} \\
            \scriptsize{Reliable} & \scriptsize{Conspiracy} & \scriptsize{Pseudosci.}
        \end{tabular}
        \\
    \end{tblr}
}

%% file: tables/generalisation_forms.tex
\begin{tabularx}{\textwidth}{Xrrrrrrrrr}
    \toprule
    \multirow{2}[2]{*}{\textbf{Generalisation Form}} & \multicolumn{3}{c}{\textbf{MCC}} & \multicolumn{3}{c}{\makecell{\textbf{F1} \\ \textbf{Reliable}}} & \multicolumn{3}{c}{\makecell{\textbf{F1} \\ \textbf{Unreliable}}} \\
    \cmidrule(lr){2-4}\cmidrule(lr){5-7}\cmidrule(lr){8-10}
    & \multicolumn{1}{c}{\textit{ID}} & \multicolumn{1}{c}{\textit{OoD}} & \multicolumn{1}{c}{$\Delta$} & \multicolumn{1}{c}{\textit{ID}} & \multicolumn{1}{c}{\textit{OoD}} & \multicolumn{1}{c}{$\Delta$} & \multicolumn{1}{c}{\textit{ID}} & \multicolumn{1}{c}{\textit{OoD}} & \multicolumn{1}{c}{$\Delta$} \\
    \midrule
    Uniform & \small{0.46}  & \small{0.46}  & \small{0.00}  & \small{0.86}  & \small{0.86}  & \small{0.00}  & \small{0.57}  & \small{0.57}  & \small{0.00} \\
    \midrule
    Time & \small{0.46}  & \small{0.33}  & -\small{0.13} & \multicolumn{6}{c}{N/A} \\
    Event & \small{0.43}  & \small{0.46}  & \small{0.03}  & \small{0.87}  & \small{0.86}  & -\small{0.01} & \small{0.52}  & \small{0.55}  & \small{0.03} \\
    Topic & \small{0.46}  & \small{0.38}  & -\small{0.08} & \small{0.87}  & \small{0.84}  & -\small{0.03} & \small{0.56}  & \small{0.50}  & -\small{0.06} \\
    Publisher & \small{0.48}  & \small{0.37}  & -\small{0.10} & \small{0.87}  & \small{0.84}  & -\small{0.03} & \small{0.58}  & \small{0.53}  & -\small{0.05} \\
    Political Bias &       &       &       &       &       &       &       &       &  \\
        \hspace{2em} \small{Left}  & \small{0.49}  & \small{0.30}  & -\small{0.19} & \small{0.85}  & \small{0.87}  & \small{0.02}  & \small{0.61}  & \small{0.38}  & -\small{0.23} \\
        \hspace{2em} \small{Right} & \small{0.56}  & \small{0.19}  & -\small{0.37} & \small{0.95}  & \small{0.60}  & -\small{0.34} & \small{0.58}  & \small{0.26}  & -\small{0.32} \\
    Misinformation Type &       &       &       &       &       &       &       &       &  \\
          \hspace{2em} \small{Consp.-PSci.} & \small{0.43}  & \small{0.42}  & -\small{0.01} & \small{0.87}  & \small{0.82}  & -\small{0.05} & \small{0.53}  & \small{0.53}  & \small{0.01} \\
          \hspace{2em} \small{Questionable} & \small{0.43}  & \small{0.23}  & -\small{0.20} & \small{0.94}  & \small{0.62}  & -\small{0.33} & \small{0.41}  & \small{0.25}  & -\small{0.16} \\
    \bottomrule
\end{tabularx}

%% file: tables/uniform_across_years.tex
\begin{tblr}{
    colspec={cccccccc},
    columns = {colsep=2pt},
    cell{3}{3} = {bg = lightgray},
    cell{4}{4} = {bg = lightgray},
    cell{5}{5} = {bg = lightgray},
    cell{6}{6} = {bg = lightgray},
    cell{7}{7} = {bg = lightgray},
    cell{8}{8} = {bg = lightgray}
}
    & & \SetVline{3-8}{black, 0.1pt} \SetCell[c=6]{c}{\textbf{Eval}} \\[-0.5em]
    & & \small{2017}  & \small{2018}  & \small{2019}  & \small{2020}  & \small{2021} & \small{2022} \\
    \SetHline{3-8}{black, 0.1pt}
    \SetCell[r=6]{m}{\rotatebox[origin=c]{90}{\textbf{Train}}}
    & \small{2017} & \small{0.50}  & \small{0.43}  & \small{0.41}  & \small{0.40}  & \small{0.40}  & \small{0.38} \\
    & \small{2018} & \small{0.29}  & \small{0.42}  & \small{0.43}  & \small{0.39}  & \small{0.41}  & \small{0.37} \\
    & \small{2019} & \small{0.26}  & \small{0.38}  & \small{0.44}  & \small{0.40}  & \small{0.41}  & \small{0.40} \\
    & \small{2020} & \small{0.34}  & \small{0.39}  & \small{0.47}  & \small{0.47}  & \small{0.47}  & \small{0.45} \\
    & \small{2021} & \small{0.31}  & \small{0.37}  & \small{0.46}  & \small{0.46}  & \small{0.47}  & \small{0.45} \\
    & \small{2022} & \small{0.33}  & \small{0.38}  & \small{0.46}  & \small{0.45}  & \small{0.46}  & \small{0.46}
\end{tblr}

%% file: tables/reasoning_models.tex
\setlength{\tabcolsep}{3pt}
\begin{tabular}{lccc}
    \toprule
    \small{\textbf{Model}} & \small{\textbf{MCC}} & \small{\makecell{\textbf{F1} \\ \textbf{Reliable}}} & \small{\makecell{\textbf{F1} \\ \textbf{Unreliable}}} \\
    \midrule
    \small{Fine-Tuned}
        & \small{0.41} & \small{0.78} & \small{0.58} \\
    \midrule
    \multicolumn{4}{l}{\small{Gemini 2.5 flash lite}} \\
        & \small{0.46} & \small{0.70} & \small{0.75} \\
    \multicolumn{4}{l}{\small{DeepSeek Reasoner}} \\
        & \small{0.52} & \small{0.77} & \small{0.76} \\
    \bottomrule
\end{tabular}

%% file: tables/uniform_predictions_emms.tex

\begin{tblr}{
  colspec={rccc},
  column{2-4} = {mode=math},
}
  & \SetVline{2-5}{black, 0.1pt} \textbf{Left} & \textbf{Center} & \textbf{Right} \\
  \SetHline{2-4}{black, 0.1pt}
  \textbf{R}       & 86.22\% & 92.90\%   & 79.03\% \\
  \textbf{Q}       & 46.13\% & 46.11\%   & 30.92\% \\
  \textbf{C}   & 32.47\% & 79.41\%   & 31.20\% \\
  \textbf{S}       &  7.10\% & \text{--} & 14.11\%
\end{tblr}

%% file: tables/datasets_survey.tex
\begin{longtblr}[
    theme = misinfodatasets,
    caption = {A long table with an (inexhaustive) sampling of misinformation datasets. Each row provides a single dataset, with name and citation, along with the labelling granularity (see Sec. \ref{sec:survey_misinformation_datasets}), the dataset size (where units `k' and `M' denote thousands and millions, respectively), and a short description of how the dataset authors generated misinformation labels.},
    label = {tab:misinfo_datasets},
    note{a} = {\href{https://www.mturk.com/}{MechanicalTurk}: crowdsourced lay volunteers},
    note{b} = {Weibo Community Management Center: credible Weibo users can report posts},
    note{c} = {\href{https://www.buzzfeed.com/about}{Buzzfeed}: a digital media company},
    note{d} = {\href{https://www.snopes.com/about/}{Snopes}: expert journalist website for debunking misinformation},
    note{e} = {\href{https://www.politifact.com/article/2018/feb/12/principles-truth-o-meter-politifacts-methodology-i/}{PolitiFact}: expert journalist website for fact checking politicians},
    note{f} = {GossipCop: a defunct website dedicated to fact-checking celebrity rumors},
    note{g} = {HealthNewsReviews: a defunct website dedicated to reviewing medical claims},
    note{h} = {\href{https://www.poynter.org/about/}{Poynter}: a global, non-profit organization with annotations from partnered organizations},
    note{i} = {\href{https://toolbox.google.com/factcheck/apis}{Google Fact Check Tool}: a unified API for fact-checking annotations}
]{
    colspec = {
        >{\centering\arraybackslash}p{0.250\textwidth}
        >{\centering\arraybackslash}p{0.100\textwidth}
        >{\centering\arraybackslash}p{0.050\textwidth}
        X
    },
    width = 1.0\linewidth,
    rowhead = 1,
    rowfoot = 0,
    hlines = {0.25pt, solid},
    hline{1, 2, Z} = {1pt, solid},
}
    \textbf{Dataset} & \textbf{Label} & \textbf{Size} & \textbf{Description} \\

    \misinfotblDataset{Lie Detector}{mihalceaLieDetectorExplorations2009} & \misinfotblLabel{Claim} & \misinfotblSize{0.3k} & \misinfotblDescription{MTurkers\TblrNote{a}\quad produce short arguments that align and oppose their stance on various topics} \\

    \misinfotblDataset{CREDBANK}{mitraCREDBANKLargeScaleSocial2015} & \misinfotblLabel{Claim} & \misinfotblSize{60M} & \misinfotblDescription{Many MTurkers\TblrNote{a}\quad annotate tweets for veracity and verifiability, with the majority annotation becoming the label} \\

    \misinfotblDataset{Weibo15}{maDetectingRumorsMicroblogs2016} & \misinfotblLabel{Article} & \misinfotblSize{5k} & \misinfotblDescription{The authors scraped user nominated misinformation articles from the Sina Weibo Community Management center\TblrNote{b}. Unannoted posts were included as factual posts} \\

    \misinfotblDataset{MediaEval15}{boididouVerifyingMultimediaUse2015} & \misinfotblLabel{Claim} & \misinfotblSize{12k} & \misinfotblDescription{The authors generated a list of events which were verified as true or false, and a collection of tweets discussing these events. The tweets were manually verified} \\

    \misinfotblDataset{BuzzFeed-Webis}{potthastStylometricInquiryHyperpartisan2018} & \misinfotblLabel{Article} & \misinfotblSize{1.6k} & \misinfotblDescription{Articles from a small set of sources were manually rated between mostly true or mostly false by expert journalists from BuzzFeed\TblrNote{c}} \\

    \misinfotblDataset{TSHP-17}{rashkinTruthVaryingShades2017} & \misinfotblLabel{Publisher} & \misinfotblSize{70k} & \misinfotblDescription{Trusted news articles were sampled from the Gigaword News corpus, whereas unreliable news was sampled from specific publishers.} \\

    \misinfotblDataset{Kaggle Fake News}{risdalGettingRealFake2016} & \misinfotblLabel{Publisher} & \misinfotblSize{13k} & \misinfotblDescription{The authors scraped articles from unreliable sources using a third-party tool. No reliable articles were included} \\

    \misinfotblDataset{Allcott \& Gentzkow}{allcottSocialMediaFake2017} & \misinfotblLabel{Article} & \misinfotblSize{0.2k} & \misinfotblDescription{Verified fake news articles were scraped from Snopes\TblrNote{d}, PolitiFact\TblrNote{e} and BuzzFeed\TblrNote{c}. No reliable articles were included} \\

    \misinfotblDataset{PHEME}{zubiagaExploitingContextRumour2017} & \misinfotblLabel{Claim} & \misinfotblSize{5.8k} & \misinfotblDescription{Tweets related to 5 mainstream events were manually annotated as unverified rumour or verified} \\

    \misinfotblDataset{Liar}{wangLiarLiarPants2017} & \misinfotblLabel{Claim} & \misinfotblSize{13k} & \misinfotblDescription{Short snippets from famous politicians scraped from the PolitiFact\TblrNote{e} API} \\

    \misinfotblDataset{Weibo17}{jinMultimodalFusionRecurrent2017} & \misinfotblLabel{Article, Publisher} & \misinfotblSize{10k} & \misinfotblDescription{They take posts reported as false from trusted users, and take articles from mainstream publishers for their factual class} \\

    \misinfotblDataset{Some Like it Hoax}{tacchiniItHoaxAutomated2017} & \misinfotblLabel{Publisher} & \misinfotblSize{15.5k} & \misinfotblDescription{Articles were scraped from Facebook groups dedicated to sharing scientific or pseudo-scientific articles} \\

    \misinfotblDataset{Fake vs Satire}{golbeckFakeNewsVs2018} & \misinfotblLabel{Publisher} & \misinfotblSize{0.5k} & \misinfotblDescription{Articles were sampled from identified satire or fake news sites. The authors constrained the number of articles per publisher to ensure a diverse publisher set. All ambiguous cases were removed} \\

    \misinfotblDataset{FakeNewsAMT}{perez-rosasAutomaticDetectionFake2018} & \misinfotblLabel{Article} & \misinfotblSize{0.5k} & \misinfotblDescription{A small set of manually verified articles were taken from mainstream publishers, and minimally edited by MTurkers\TblrNote{a}\quad to produce misinformation} \\

    \misinfotblDataset{Web Dataset Celebrity}{perez-rosasAutomaticDetectionFake2018} & \misinfotblLabel{Article} & \misinfotblSize{0.5k} & \misinfotblDescription{To complement FakeNewsAMT, the authors collect articles from rumour and tabloid publications, and manually verify articles using sites like GossipCop\TblrNote{f}} \\

    \misinfotblDataset{\makecell{FakeNewsNet \\ PolitiFact}}{shuFakeNewsNetDataRepository2019} & \misinfotblLabel{Claim, Article} & \misinfotblSize{23k} & \misinfotblDescription{The authors label an articles based on a claim made within, where the claim is labelled by PolitiFact\TblrNote{e}} \\

    \misinfotblDataset{\makecell{FakeNewsNet \\ GossipCop}}{shuFakeNewsNetDataRepository2019} & \misinfotblLabel{Article, Publisher} & \misinfotblSize{23k} & \misinfotblDescription{Unverified rumour articles were taken from GossipCop\TblrNote{f}, with verified rumours coming from a few mainstream publishers} \\

    \misinfotblDataset{FakeNewsCorpus}{pathakBREAKINGPresentingFake2019} & \misinfotblLabel{Article, Publisher} & \misinfotblSize{0.7k} & \misinfotblDescription{\textasciitilde700 articles were taken from questionable source publishers, and used as misinformation, and 26 expert labelled factual news articles. Satire and unverifiable news were explicitly excluded.} \\

    \misinfotblDataset{QProp}{barron-cedenoProppyOrganizingNews2019} & \misinfotblLabel{Publisher} & \misinfotblSize{51k} & \misinfotblDescription{Uses MBFC to assign articles the label of their publisher. They manage to sample from 104 different sources, although only include 10 progandistic sources.} \\

    \misinfotblDataset{Przybyła Credibility}{przybylaCapturingStyleFake2020} & \misinfotblLabel{Publisher} & \misinfotblSize{100k} & \misinfotblDescription{Scrapes articles from websites classified as non-credible by PolitiFact\TblrNote{e}. The authors specifically evaluate publishers as credible or non-credible, as opposed to fake or factual news.} \\

    \misinfotblDataset{FakeHealth}{daiGingerCannotCure2020} & \misinfotblLabel{Article} & \misinfotblSize{2.3k} & \misinfotblDescription{Both variants of the dataset (HealthStory and HealthRelease) include text manually verified by experts from HealthNewsReview.org\TblrNote{g} on the \textit{credibility} of the information provided} \\

    \misinfotblDataset{MM-COVID}{liMMCOVIDMultilingualMultimodal2020} & \misinfotblLabel{Article, Publisher} & \misinfotblSize{4.2k} & \misinfotblDescription{Articles with manual labels were collected from Snopes\TblrNote{d} and Poynter\TblrNote{h}, and to complement reliable articles, they sample from mainstream media sources} \\

    \misinfotblDataset{FakeCovid}{shahiFakeCovidMultilingualCrossdomain2020} & \misinfotblLabel{Article} & \misinfotblSize{5.2k} & \misinfotblDescription{Specifically COVID articles with labels from Snopes\TblrNote{d} and Poynter\TblrNote{h} were collected. The authors make sure to include labels from 92 separate organizations across 105 countries} \\

    \misinfotblDataset{WeChat}{wangWeakSupervisionFake2020} & \misinfotblLabel{Article} & \misinfotblSize{4k} & \misinfotblDescription{The authors collected articles flagged by WeChat users. A small subset was annotated by experts, while a larger subset was unannotated, meant for unsupervised training} \\

    \misinfotblDataset{CoAID}{cuiCoAIDCOVID19Healthcare2020a} & \misinfotblLabel{Claim, Article, Publisher} & \misinfotblSize{3.7k} & \misinfotblDescription{Misinformation articles about the COVID19 pandemic were scraped directly from various fact-checking sources. Factual articles were scraped from 9 reliable publishers. Claims were scraped from official government sites} \\

    \misinfotblDataset{MuMIN}{nielsenMuMiNLargeScaleMultilingual2022} & \misinfotblLabel{Claim} & \misinfotblSize{13k} & \misinfotblDescription{The authors collected a set of 115 fact checking organisation from the Google Fact Check Tool\TblrNote{i} API, and then collected all fact-checked claims from these organisations. They use a separate classifier to collate different labelling schemas} \\

    \misinfotblDataset{PolitiFact-Oslo}{poldverePolitiFactOsloCorpusNew2023} & \misinfotblLabel{Claim, Article} & \misinfotblSize{2.7k} & \misinfotblDescription{Claims were extracted from PolitiFact\TblrNote{e}, and the post or article from which the claim originated was manually extracted. The authors specifically highlight the importance of publisher-level metadata} \\

    \misinfotblDataset{MCFEND}{liMCFENDMultisourceBenchmark2024} & \misinfotblLabel{Article} & \misinfotblSize{24K} & \misinfotblDescription{Articles annotated by various fact-checking organisations around the world were collected, and manually mapped to a single annotation schema.} \\
\end{longtblr}

%% file: tables/dataset_sizes.tex
\begin{tabularx}{\textwidth}{Xccccccc}
    \toprule
    \textbf{Year}  & \textbf{2017}  & \textbf{2018}  & \textbf{2019}  & \textbf{2020}  & \textbf{2021}  & \textbf{2022}  & \textbf{Total} \\
    \midrule 
    Original
        & \small{0.14}  & \small{0.70}  & \small{1.12}  & \small{1.78}  & \small{1.86}  & \small{1.78}  & \small{7.24} \\
    \midrule \\[-1.50em]
    \multirow{1.5}[0]{*}{De-aggregation \& Labelling}
        & \small{0.13}  & \small{0.61}  & \small{0.96}  & \small{1.62}  & \small{1.71}  & \small{1.66}  & \small{6.69}  \\[-0.25em]
        & \small{-1\%}  & \small{-9\%}  & \small{-12\%} & \small{-5\%}  & \small{-3\%}  & \small{-3\%}  & \small{-8\%}  \\
    \multirow{1.5}[0]{*}{Exact Deduplication}
        & \small{0.12}  & \small{0.55}  & \small{0.86}  & \small{1.34}  & \small{1.39}  & \small{1.32}  & \small{5.58}  \\[-0.25em]
        & \small{-6\%}  & \small{-11\%} & \small{-12\%} & \small{-18\%} & \small{-20\%} & \small{-21\%} & \small{-23\%} \\
    \multirow{1.5}[0]{*}{Cleaning}
        & \small{0.12}  & \small{0.53}  & \small{0.71}  & \small{1.12}  & \small{1.16}  & \small{1.11}  & \small{4.74}  \\[-0.25em]
        & \small{-3\%}  & \small{-4\%}  & \small{-17\%} & \small{-16\%} & \small{-17\%} & \small{-16\%} & \small{-35\%} \\
    \multirow{1.5}[0]{*}{Near Deduplication}
        & \small{0.11}  & \small{0.47}  & \small{0.65}  & \small{1.03}  & \small{1.06}  & \small{1.01}  & \small{4.33}  \\[-0.25em]
        & \small{-11\%} & \small{-12\%} & \small{-8\%}  & \small{-8\%}  & \small{-9\%}  & \small{-9\%}  & \small{-40\%} \\
    \multirow{1.5}[0]{*}{Language Detection}
        & \small{0.11}  & \small{0.47}  & \small{0.64}  & \small{1.02}  & \small{1.05}  & \small{0.99}  & \small{4.16}  \\[-0.25em]
        & \small{0\%}   & \small{-1\%}  & \small{-1\%}  & \small{-1\%}  & \small{-1\%}  & \small{-1\%}  & \small{-43\%} \\
    \bottomrule
\end{tabularx}

%% file: tables/banned_publishers.tex
\begin{tabular}{lr}
    \toprule
    \textbf{Publisher} & \textbf{\# Articles} \\
    \midrule
    drudgereport       & 92 186      \\
    bonginoreport      & 17 895      \\
    whatreallyhappened & 57 431      \\
    thelibertydaily    & 6 346       \\
    yahoonews          & 52 797      \\
    theduran           & 10 722      \\
    \midrule
    Total              & 237 377     \\
    \bottomrule
\end{tabular}

%% file: tables/banned_domains.tex
\begin{tabular}{lr}
    \toprule
    \textbf{Domain} & \textbf{\# Articles} \\
    \midrule
    soundcloud.com  & 106        \\
    youtu.be        & 4 009       \\
    apps.apple.com  & 414        \\
    amazon.com      & 112        \\
    facebook        & 108        \\
    amazon          & 113        \\
    youtube         & 2 902       \\
    play.google.com & 421        \\
    twitter         & 274        \\
    instagram       & 15         \\
    reddit          & 4          \\
    dnyuz.com       & 1 900       \\
    \midrule
    Total           & 10 378      \\
    \bottomrule
\end{tabular}

%% file: tables/frequent_duplicates.tex
\begin{tabular}{ll}
    \toprule
    \textbf{Type}                 & \textbf{Sample Text} \\
    \midrule
    Banners                       & Click for more article by Guest \ldots \\
    Previews                      & To read the full blog, please check out the complete post \ldots \\
    Prayers                       & \makecell[lc]{Our Father, who art in heaven, hallowed be thy Name \ldots} \\
    Podcast Descriptions & Don't forget to tune in to \ldots \\
    Error Messages                & 403 Forbidden nginx \\
    \bottomrule
\end{tabular}

%% file: tables/rules.tex
\begin{tabularx}{\textwidth}{Xc}
    \toprule
    \textbf{Rule}                                                                 & \textbf{VALUE} \\
    \midrule
    Articles must contain at least \$\{VALUE\} tokens                             & 16    \\
    Articles must not contain more than \$\{VALUE\} tokens                        & 4096  \\
    The article must have a title                                                 & -     \\
    The article must be at least \$\{VALUE\} times longer than its title          & 3     \\
    The article must have a mean token length greater than \$\{VALUE\}            & 2     \\
    The article must have a mean token length less than \$\{VALUE\}               & 10    \\
    The article must have at most \$\{VALUE\}\% copyright tokens                  & 20    \\
    \bottomrule
\end{tabularx}

%% file: tables/publisher_changed_labels.tex
\begin{tblr}{
    cells   = {font=\footnotesize\selectfont},
    colspec={lcccl},
    columns = {colsep=4pt},
    hline{1}  = {1-5}{1pt, solid},
    hline{2}  = {1-5}{1pt, solid},
    hline{22} = {1-5}{1pt, solid},
}
    \textbf{Publisher Name} & \textbf{Previous Label} & \textbf{New Label} & \textbf{Year} & \textbf{Reason} \\
    Big League Politics & Right         & Questionable	& 2019 & Failed fact checks \\
    Daily Wire          & Questionable  & Right         & 2021 & Correcting fact checks \\
    Ecowatch            & Pseudoscience & Left	        & 2020 & Editorial improvements \\
    End Time Headlines  & Conspiracy    & Right-Center	& 2022 & Improved publishing \\
    FoxNews             & Right         & Questionable	& 2021 & Failed fact checks, state propaganda \\
    Just the News       & Right         & Questionable	& 2021 & Failed fact checks \\
    Newsmax             & Right         & Questionable	& 2020 & Failed fact checks \\
    One America News    & Right         & Questionable	& 2020 & Failed fact checks \\
    Pravda Report       & Right         & Questionable	& 2020 & \\
    Russia Insider      & Right-Center  & Questionable	& 2020 & \\
    Strategic Culture Foundation & Right-Center & Questionable & 2019 & Propaganda, conspiracy theories \\
    The Drudge Report   & Questionable  & Right-Center	& 2020 & Improved standards \\
    The Political Insider & Right       & Questionable	& 2019 & \\
    Townhall            & Right         & Questionable	& 2020 & Failed fact checks \\
    Truth Theory        & Pseudoscience & Left	        & 2020 & Change in direction \\
    Turning Point USA   & Right         & Questionable	& 2019 & Failed fact checks \\
    Washington Times    & Right-Center  & Questionable	& 2020 & Failed fact checks \\
    Western Journal     & Right         & Questionable	& 2020 & Failed fact checks \\
    WhatFinger          & Right         & Questionable	& 2022 & Conspiracy theories \\
    Wings over Scotland & Left-Center   & Questionable  & 2021 & Conspiracies, hate speech \\
\end{tblr}

%% file: tables/dataset_counts.tex
\setlength{\tabcolsep}{4pt}
\begin{tabular}{llcccccc}
    \toprule
    \multicolumn{2}{c}{\textbf{Statistic}} & \textbf{2017} & \textbf{2018} & \textbf{2019} & \textbf{2020} & \textbf{2021} & \textbf{2022} \\
    \midrule
    \multicolumn{2}{l}{Article Counts} & 0.10M & 0.46M & 0.59M & 1.02M & 1.04M & 0.99M \\
    \midrule
    \multirow{2}{*}{Publishers}
        & Reliable & 43 & 99 & 134 & 189 & 183 & 184 \\
        & Unreliable & 49 & 60 & 100 & 249 & 219 & 201 \\
    \midrule
    \multirow{4}{*}{\makecell[l]{Truncated \\ Token \\ Counts}}
        & Mean & 398.53 & 383.59 & 430.22 & 432.87 & 428.15 & 435.90 \\
        & Q25 & 299 & 266 & 358 & 369 & 366 & 383 \\
        & Q50 & 488 & 448 & 512 & 512 & 512 & 512 \\
        & Q75 & 512 & 512 & 512 & 512 & 512 & 512 \\
    \midrule
    \multirow{4}{*}{\makecell[l]{Event \\ Clustering}}
        & \# Events & 2674 & 6288 & 7718 & 7931 & 8758 & 8336  \\
        & \small{Mean Event Size} & 38.67 & 73.38 & 76.87 & 128.49 & 118.24 & 119.29 \\
        & \small{Smallest Topic Size} & 2.7K & 20.5K & 24.6K & 27.7K & 51.3K & 48.4K \\
        & \small{Largest Topic Size} & 38.3K & 94.1K & 113.1K & 192.3K & 211.4K & 209.4K \\
    \midrule
    \multirow{2}{*}{\makecell[l]{Label \\ Proportion}}
        & \reliable   & 61.39\% & 72.82\% & 75.91\% & 70.73\% & 71.36\% & 72.00\% \\
        & \unreliable & 38.61\% & 27.18\% & 24.09\% & 29.27\% & 28.64\% & 28.00\% \\
    \midrule
    \multirow{6}{*}{\makecell[l]{\reliable}}
        & Left & \small{17.91\%} & \small{13.32\%} & \small{11.13\%} & \small{9.01\%} & \small{8.76\%} & \small{9.88\%} \\
        & Left-Center & \small{24.09\%} & \small{29.62\%} & \small{35.19\%} & \small{36.50\%} & \small{34.53\%} & \small{35.01\%} \\
        & Least Biased & \small{0.93\%} & \small{2.18\%} & \small{4.38\%} & \small{4.91\%} & \small{5.13\%} & \small{3.97\%} \\
        & Right-Center & \small{4.34\%} & \small{12.66\%} & \small{9.81\%} & \small{8.70\%} & \small{10.38\%} & \small{9.52\%} \\
        & Right & \small{14.12\%} & \small{15.05\%} & \small{15.41\%} & \small{11.40\%} & \small{12.03\%} & \small{13.08\%} \\
        & Pro-Science & & & & \small{0.21\%} & \small{0.53\%} & \small{0.55\%} \\
    \midrule
    \multirow{3}{*}{\makecell[l]{\unreliable}}
        & Satire & \small{1.70\%} & \small{0.99\%} & \small{0.82\%} & \small{0.66\%} & \small{0.48\%} & \small{0.42\%} \\
        & \small{Questionable Source} & \small{27.83\%} & \small{20.14\%} & \small{17.34\%} & \small{19.41\%} & \small{19.00\%} & \small{18.45\%} \\
        & \small{Consp.-Pseudosci.} & \small{9.09\%} & \small{6.05\%} & \small{5.93\%} & \small{9.20\%} & \small{9.17\%} & \small{9.13\%} \\
    \bottomrule
\end{tabular}

%% file: tables/topic_example.tex
\begin{tblr}{
    colspec={clccccc},
    hline{1, 3, 13} = {1pt,solid},
    hline{2} = {5,6}{0.5pt, solid},
    hline{9} = {0.5pt, dashed},
    cell{1}{1} = {r=2,c=1}{c},
    cell{1}{2} = {r=2,c=1}{c},
    cell{1}{3} = {r=2,c=2}{c},
    cell{1}{5} = {r=1,c=2}{c},
    cell{1}{7} = {r=2,c=1}{c},
}
\textbf{\#} & \textbf{Description} & \textbf{\# Articles} & & \textbf{Labels} & & \textbf{ID/OoD} \\
& & & & \reliable & \unreliable & \\
7 & \makecell[l]{US Federal Elections}      & 192k & 18.8\% & 64.11\% & 35.89\% & ID \\ 
9 & \makecell[l]{International Affairs}     & 161k & 15.8\% & 75.80\% & 24.20\% & ID \\ 
8 & \makecell[l]{Health \& COVID}           & 149k & 14.6\% & 70.85\% & 29.16\% & ID \\ 
4 & \makecell[l]{Entertainment \& Sports}   & 112k & 11.0\% & 86.30\% & 13.70\% & ID \\ 
5 & \makecell[l]{Economy \& Social Issues}  & 112k & 10.9\% & 63.50\% & 36.50\% & ID \\ 
6 & \makecell[l]{Science \& Technology}     &  89k &  8.7\% & 71.68\% & 28.31\% & ID \\ 
3 & \makecell[l]{US Local Politics}         &  66k &  6.5\% & 68.85\% & 31.15\% & OoD \\ 
2 & \makecell[l]{Crime \& Justice}          &  64k &  6.2\% & 68.93\% & 31.07\% & OoD \\ 
0 & Conflict                                &  45k &  4.4\% & 69.37\% & 30.63\% & OoD \\ 
1 & \makecell[l]{Biden Administration}      &  28k &  2.7\% & 60.43\% & 39.57\% & OoD \\
\end{tblr}

%% file: tables/publisher_overlap.tex
\begin{tblr}{
    colspec={ccccccc},
    columns = {colsep=4pt},
    cell{3}{3} = {bg = lightgray},
    cell{4}{4} = {bg = lightgray},
    cell{5}{5} = {bg = lightgray},
    cell{6}{6} = {bg = lightgray},
    cell{7}{7} = {bg = lightgray},
    cell{8}{8} = {bg = lightgray}
}
    & & \SetVline{3-8}{black, 0.1pt} \\[-0.5em]
    & & \small{2017}  & \small{2018}  & \small{2019}  & \small{2020}  & \small{2021} & \small{2022} \\
    \SetHline{3-8}{black, 0.1pt}
    & \small{2017} & \small{1.00} & \small{0.43} & \small{0.31} & \small{0.17} & \small{0.16} & \small{0.16} \\
    & \small{2018} & \small{0.43} & \small{1.00} & \small{0.62} & \small{0.34} & \small{0.33} & \small{0.32} \\
    & \small{2019} & \small{0.31} & \small{0.62} & \small{1.00} & \small{0.52} & \small{0.50} & \small{0.49} \\
    & \small{2020} & \small{0.17} & \small{0.34} & \small{0.52} & \small{1.00} & \small{0.86} & \small{0.80} \\
    & \small{2021} & \small{0.16} & \small{0.33} & \small{0.50} & \small{0.86} & \small{1.00} & \small{0.89} \\
    & \small{2022} & \small{0.16} & \small{0.32} & \small{0.49} & \small{0.80} & \small{0.89} & \small{1.00}
\end{tblr}

%% file: tables/publisher_topic_overlap.tex
\begin{tabular}{cccccc}
    \toprule
    \multirow{2}{*}{\textbf{Year}} & \multirow{2}{*}{\textbf{IoU}} & \multirow{2}{*}{\textbf{\makecell{Random \\ IoU}}} & \multirow{2}{*}{\textbf{\makecell{Execess \\ IoU}}} & \multicolumn{2}{c}{\textbf{Test Set Sizes}} \\
         &        &        &        & \textit{Publisher} & \textit{Topic} \\
    \midrule
    2017 & 0.1183 & 0.1084 &  9.07\% &  19764 &  20709 \\
    2018 & 0.1227 & 0.1162 &  5.62\% & 109043 &  85832 \\
    2019 & 0.1210 & 0.1133 &  6.80\% & 133465 & 110168 \\
    2020 & 0.1386 & 0.1116 & 24.24\% & 203224 & 205873 \\
    2021 & 0.1317 & 0.1055 & 24.82\% & 192315 & 203311 \\
    2022 & 0.1318 & 0.0992 & 32.93\% & 172530 & 186986 \\
    \bottomrule
\end{tabular}

%% file: tables/hyperparameters.tex
\begin{tabular}{lc}
    \toprule
        \textbf{Hyperparameter} & \textbf{Value} \\
    \midrule
        \texttt{seed} & 942 \\
        \texttt{max\_length} & 512 \\
        \texttt{val\_prop} & 0.1 \\
        \texttt{batchsize} & 64 \\
        \texttt{pooler\_dropout} & 0.1 \\
        \multicolumn{2}{l}{\texttt{lr}} \\
            \quad \texttt{embeddings} & 5.0e-7 \\
            \quad \texttt{pooler} & 5.0e-5 \\
            \quad \texttt{classifier} & 1.0e-3 \\
        \texttt{weight\_decay} & 0.1 \\
        \multicolumn{2}{l}{\texttt{lr\_scheduler}} \\
            \quad \texttt{type} & `polynomial' \\
            \quad \texttt{power} & 3.163 \\
        \texttt{max\_steps} & 3.0e+6 \\
        \texttt{warmup\_ratio} & 0.0213 \\
        \texttt{eval\_prop} & 0.05 \\
        \texttt{patience} & 2 \\
    \bottomrule
\end{tabular}

%% file: tables/subjectivity_annotation_confusion_matrix.tex
\begin{tblr}{
    colspec={ccccccc},
    columns = {colsep=1pt},
    vline{3} = {3-8}{0.5pt, solid},
    hline{3} = {3-8}{0.5pt, solid},
    cell{3}{3} = {bg = lightgray},
    cell{4}{4} = {bg = lightgray},
    cell{5}{5} = {bg = lightgray},
    cell{6}{6} = {bg = lightgray},
    cell{7}{7} = {bg = lightgray},
}
    & & \footnotesize{E. Obj.}  & \footnotesize{M. Obj.}  & \footnotesize{Mixed}  & \footnotesize{M. Subj.}  & \footnotesize{E. Subj.} \\
    & \footnotesize{E. Obj.} & \footnotesize{5.71\%} & \footnotesize{5.13\%} & \footnotesize{0.39\%} & \footnotesize{0.01\%} & \\
    & \footnotesize{M. Obj.} & \footnotesize{1.80\%} & \footnotesize{12.30\%} & \footnotesize{3.76\%} & \footnotesize{0.97\%} & \\
    & \footnotesize{Mixed} & & \footnotesize{3.37\%} & \footnotesize{15.19\%} & \footnotesize{7.62\%} & \footnotesize{0.49\%} \\
    & \footnotesize{M. Subj.} & \footnotesize{0.05\%} & \footnotesize{0.39\%} & \footnotesize{4.35\%} & \footnotesize{14.70\%} & \footnotesize{4.00\%} \\
    & \footnotesize{E. Subj.} & & & \footnotesize{0.83\%} & \footnotesize{5.08\%} & \footnotesize{13.77\%}
\end{tblr}

%% file: tables/dataset_effect_counts.tex
\begin{tabular}{cccc}
    \toprule
    \textbf{Quantity} & \textbf{Top 1} & \textbf{Top 2} & \textbf{Top 3} \\
    \midrule
     Articles & 28\% & 43\% & 55\% \\
     Publisher & 11\% & 20\% & 30\% \\
     \bottomrule
\end{tabular}

%% file: tables/determinants_table.tex
\small
\setlength{\tabcolsep}{1pt}
\begin{tabular}{clccccccc}
\toprule
\textbf{Group} & \multicolumn{1}{c}{\textbf{Parameter}} & \textbf{Range} & \boldmath{}\textbf{$\beta$}\unboldmath{} & \boldmath{}\textbf{$\exp(\beta)$}\unboldmath{} & \textbf{Std. Err.} & \textbf{p} & \multicolumn{2}{c}{\textbf{95\% CI}} \\
\midrule
\textit{Intercept}
        & Intercept & 1     & -0.85 & 0.43  & 0.03  & 0.00  & -0.91 & -0.79 \\
\midrule
\multirow{1.5}[2]{*}{\textit{Generic}}
        & $\log_{10}(\text{train count})$ & $[0, \infty]$ & 0.65  & 1.91  & 0.01  & 0.00  & 0.64  & 0.66 \\
        & Foreign & 0..1  & 1.07  & 2.91  & 0.01  & 0.00  & 1.06  & 1.08 \\
\midrule
\multirow{4.5}[2]{*}{\textit{Year}}
        & 2018  & 0..1  & 0.07  & 1.08  & 0.02  & 0.00  & 0.04  & 0.11 \\
        & 2019  & 0..1  & 0.14  & 1.15  & 0.02  & 0.00  & 0.10  & 0.17 \\
        & 2020  & 0..1  & -0.16 & 0.85  & 0.02  & 0.00  & -0.19 & -0.13 \\
        & 2021  & 0..1  & -0.06 & 0.94  & 0.02  & 0.00  & -0.09 & -0.03 \\
        & 2022  & 0..1  & 0.00  & 1.00  & 0.02  & 0.85  & -0.03 & 0.04 \\
\midrule
\multirow{1.5}[2]{*}{\makecell{\textit{Political} \\ \textit{Bias}}}
        & Left  & 0..1  & -0.39 & 0.68  & 0.01  & 0.00  & -0.41 & -0.37 \\
        & Right & 0..1  & -0.64 & 0.53  & 0.01  & 0.00  & -0.66 & -0.62 \\
\midrule
\multirow{2.5}[2]{*}{\textit{Label}}
        & Conspiracy-PseudoScience & 0..1  & -0.32 & 0.73  & 0.22  & 0.15  & -0.76 & 0.11 \\
        & Questionable Source & 0..1  & -2.40 & 0.09  & 0.02  & 0.00  & -2.44 & -2.37 \\
        & Satire & 0..1  & -1.58 & 0.21  & 0.18  & 0.00  & -1.94 & -1.23 \\
\midrule
\multirow{5.5}[2]{*}{\textit{Interactions}}
        & \small{Left : Conspiracy-PseudoScience} & 0..1  & -1.65 & 0.19  & 0.19  & 0.00  & -2.02 & -1.28 \\
        & \small{Left : Questionable Source} & 0..1  & 1.54  & 4.69  & 0.03  & 0.00  & 1.48  & 1.61 \\
        & \small{Left : Satire} & 0..1  & -2.15 & 0.12  & 0.20  & 0.00  & -2.54 & -1.75 \\
        & \small{Right : Conspiracy-PseudoScience} & 0..1  & -1.70 & 0.18  & 0.19  & 0.00  & -2.06 & -1.33 \\
        & \small{Right : Questionable Source} & 0..1  & 1.20  & 3.31  & 0.02  & 0.00  & 1.17  & 1.23 \\
        & \small{Right : Satire} & 0..1  & 0.56  & 1.75  & 0.35  & 0.11  & -0.12 & 1.25 \\
\midrule
\multirow{1.5}[2]{*}{\textit{Strength}}
        & Conspiracy & 1..5  & 0.19  & 1.21  & 0.02  & 0.00  & 0.15  & 0.22 \\
        & PseudoScience & 1..5  & 0.17  & 1.18  & 0.01  & 0.00  & 0.14  & 0.19 \\
\midrule
\multirow{2.5}[2]{*}{\makecell{\textit{Factuality} \\ \textit{Score}}}
        & \small{Factuality : Reliable} & 1..5  & 0.35  & 1.42  & 0.00  & 0.00  & 0.34  & 0.36 \\
        & \small{Factuality : Conspiracy-PseudoScience} & 1..5  & -0.09 & 0.92  & 0.03  & 0.00  & -0.14 & -0.04 \\
        & \small{Factuality : Questionable Source} & 1..5  & -0.13 & 0.88  & 0.01  & 0.00  & -0.14 & -0.11 \\
\bottomrule
\end{tabular}%

%% file: tables/determinants_table_topic.tex
\small
\setlength{\tabcolsep}{1pt}
\begin{tabular}{clccccccc}
\toprule
\textbf{Group} & \multicolumn{1}{c}{\textbf{Parameter}} & \textbf{Range} & \boldmath{}\textbf{$\beta$}\unboldmath{} & \boldmath{}\textbf{$\exp(\beta)$}\unboldmath{} & \textbf{Std. Err.} & \textbf{p} & \multicolumn{2}{c}{\textbf{95\% CI}} \\
\midrule
\textit{Intercept} &
        Intercept & 1     & -0.49 & 0.61  & 0.05  & 0.00  & -0.60 & -0.38 \\
\midrule
\textit{Distance} &
        Topic Distance & [0, 1] & 0.15  & 1.16  & 0.03 & 0.00 & 0.09 & 0.21 \\
\midrule
\multirow{1.5}[2]{*}{\textit{Generic}}
        & $\log_{10}(\text{train count})$ & [0, $\infty$] & 0.56  & 1.75  & 0.01  & 0.00  & 0.54  & 0.58 \\
        & Foreign & 0..1  & 0.88  & 2.42  & 0.01  & 0.00  & 0.86  & 0.90 \\
\midrule
\multirow{4.5}[2]{*}{\textit{Year}}
        & 2018  & 0..1  & 0.23  & 1.26  & 0.03  & 0.00  & 0.17  & 0.29 \\
        & 2019  & 0..1  & 0.30  & 1.35  & 0.03  & 0.00  & 0.24  & 0.36 \\
        & 2020  & 0..1  & -0.18 & 0.83  & 0.03  & 0.00  & -0.24 & -0.13 \\
        & 2021  & 0..1  & -0.47 & 0.62  & 0.03  & 0.00  & -0.53 & -0.41 \\
        & 2022  & 0..1  & -0.27 & 0.76  & 0.03  & 0.00  & -0.33 & -0.21 \\
\midrule
\multirow{1.5}[2]{*}{\makecell{\textit{Political} \\ \textit{Bias}}}
        & Left  & 0..1  & -2.81 & 0.06  & 0.47  & 0.00  & -3.74 & -1.88 \\
        & Right & 0..1  & -2.58 & 0.08  & 0.03  & 0.00  & -2.64 & -2.52 \\
\midrule
\multirow{2.5}[2]{*}{\textit{Label}}
        & Conspiracy-PseudoScience & 0..1  & -1.51 & 0.22  & 0.25  & 0.00  & -1.99 & -1.03 \\
        & Questionable Source & 0..1  & -0.54 & 0.58  & 0.01  & 0.00  & -0.57 & -0.51 \\
        & Satire & 0..1  & -0.73 & 0.48  & 0.02  & 0.00  & -0.76 & -0.70 \\
\midrule
\multirow{5.5}[2]{*}{\textit{Interactions}}
        & \small{Left : Conspiracy-PseudoScience} & 0..1  & -0.38 & 0.69  & 0.41  & 0.36  & -1.18 & 0.43 \\
        & \small{Left : Questionable Source} & 0..1  & 2.11  & 8.27  & 0.05  & 0.00  & 2.02  & 2.21 \\
        & \small{Left : Satire} & 0..1  & -1.83 & 0.16  & 0.31  & 0.00  & -2.44 & -1.22 \\
        & \small{Right : Conspiracy-PseudoScience} & 0..1  & -0.73 & 0.48  & 0.41  & 0.07  & -1.53 & 0.06 \\
        & \small{Right : Questionable Source} & 0..1  & 1.92  & 6.85  & 0.03  & 0.00  & 1.87  & 1.98 \\
        & \small{Right : Satire} & 0..1  & 0.32  & 1.37  & 0.45  & 0.48  & -0.56 & 1.19 \\
\midrule
\multirow{1.5}[2]{*}{\textit{Strength}}
        & Conspiracy & 1..5  & 0.42  & 1.52  & 0.04  & 0.00  & 0.35  & 0.49 \\
        & PseudoScience & 1..5  & 0.29  & 1.33  & 0.03  & 0.00  & 0.24  & 0.34 \\
\midrule
\multirow{2.5}[2]{*}{\makecell{\textit{Factuality} \\ \textit{Score}}}
        & \small{Factuality : Reliable} & 1..5  & 0.38  & 1.47  & 0.01  & 0.00  & 0.37  & 0.40 \\
        & \small{Factuality : Conspiracy-PseudoScience} & 1..5  & 0.42  & 1.53  & 0.05  & 0.00  & 0.33  & 0.52 \\
        & \small{Factuality : Questionable Source} & 1..5  & -0.44 & 0.65  & 0.01  & 0.00  & -0.46 & -0.41 \\
\bottomrule
\end{tabular}